\documentclass[blind, 3p,times,11pt,onecolumn,nonatbib]{elsarticle}

\usepackage{amssymb}
\usepackage{amsmath} 
\usepackage{algorithm}
\usepackage{algpseudocode}
\usepackage{booktabs}
\usepackage{multirow}
\usepackage{float}
\usepackage{dsfont}
\usepackage{tabularx}
\usepackage[colorlinks=true, citecolor=blue]{hyperref}
\makeatletter\let\c@author\relax\let\elsbibfont\bibfont\let\bibfont\relax\makeatother
\usepackage[ backend=biber,
style=apa,
hyperref=true,%
backref=false,%
sorting=nty,%
maxnames=99,%
isbn=false,%
block=ragged,%
dashed=false,%
maxcitenames=2
]{biblatex}
\renewcommand\bibfont{\elsbibfont}

\DeclareCiteCommand{\cite}
  {\ifentrytype{commentary}{}{\usebibmacro{prenote}}}
  {\usebibmacro{citeindex}%
   \printtext[bibhyperref]{%
     \printnames{labelname}%
     \addspace%
     \mkbibparens{\printfield{year}}}}
  {\multicitedelim}
  {\usebibmacro{postnote}}

\DeclareFieldFormat{bibhyperrefnonest}{%
  \DeclareFieldFormat{bibhyperref}{##1}%
  \bibhyperref{#1}}

\DeclareCiteCommand{\parencite}[\mkbibparens]
  {\usebibmacro{cite:init}%
   \usebibmacro{prenote}}
  {\usebibmacro{citeindex}%
   \printtext[bibhyperrefnonest]{\usebibmacro{cite}}}
  {}
  {\usebibmacro{postnote}%
   \usebibmacro{cite:post}}

\journal{International Journal of Forecasting}

\begin{document}
\begin{frontmatter}



\title{Deep Learning vs.\ Statistical Models for Multi-Horizon Price Forecasting of Second-Hand Electronics: A Systematic Benchmark}


\author[a,e]{Mateusz Buczyński}
\ead{mp.buczynski2@uw.edu.pl}
\author[a,e]{Michał Woźniak}
\ead{mj.wozniak9@uw.edu.pl}
\author[a,e]{Konrad Kaczyński}
\ead{k.kaczynski@student.uw.edu.pl}
\author[d,e]{Anna Wróblewska}
\ead{anna.wrobewska1@pw.edu.pl}
\author[e]{Sebastian Kuk}
\ead{sebastian.kuk@wesub.pl}

\affiliation[a]{organization={Faculty of Economic Sciences, University of Warsaw},
            addressline={Dluga 44/50},
            city={Warsaw},
            postcode={00-241},
            country={PL}}

\affiliation[d]{organization={Faculty of Mathematics and Information Science, Warsaw University of Technology},
            addressline={Koszykowa 75},
            city={Warsaw},
            postcode={00-662},
            country={PL}}

\affiliation[e]{organization={WeSub},
            addressline={Branickiego 15},
            city={Warsaw},
            postcode={02-972},
            country={PL}}

\begin{abstract}
Forecasting resale prices of used electronics is critical for subscription-based platforms where pricing errors translate directly into risk. Unlike structured financial markets, second-hand electronics exhibit high volatility, sparse listing histories, and non-normal price dynamics - yet no systematic time-series benchmark exists for this domain. This paper presents the first multi-horizon benchmark of statistical and deep learning forecasting models for used electronics price prediction. We use a large-scale dataset of daily price listings from Polish online marketplaces (January 2022 to March 2025, 100+ smartphone and laptop models) and evaluate eleven models across six horizons from 1 to 365 days, covering classical methods (ARIMA, ETS, Theta), recurrent and convolutional networks (LSTM, TCN), and modern deep architectures (N-BEATS, N-HiTS, TFT, PatchTST, Informer). Three complementary evaluation protocols assess trajectory fitness, one-shot endpoint accuracy, and cross-horizon transfer. N-BEATS achieves the lowest MAPE beyond 30 days, reaching 8.51\% at 365 days versus 14.94\% for the best statistical baseline - a 43\% reduction. At short horizons (1--7 days), all models converge near 0.72\% MAPE and the na\"{i}ve baseline remains competitive. A single N-BEATS model trained at 365 days generalizes to all shorter horizons, eliminating the need for horizon-specific models. N-BEATS and N-HiTS also demonstrate superior hyperparameter stability.

\end{abstract}

\begin{highlights}
\item First systematic multi-horizon benchmark for used electronics price forecasting.
\item Neural basis expansion achieves best accuracy from 30 days to one year ahead.
\item Recurrent and convolutional models degrade sharply beyond 90-day horizons.
\item A single one-year model generalizes to all shorter horizons without retraining.
\item Deep learning reduces forecasting error by 43\% versus best statistical baseline.
\end{highlights}


\end{frontmatter}



\section{Introduction}
\label{sec:intro}

Forecasting the prices of used electronic items presents unique challenges: rapid technological advancement drives non-linear depreciation, market dynamics are shaped by individual seller strategies rather than centralized exchanges, and historical data is scattered across regional classified platforms and social media groups. Unlike new products, second-hand items lack standardized pricing - their value depends on condition, brand perception, and momentary supply - making predictive accuracy genuinely difficult to achieve in practice.

The problem is made commercially acute by the growth of subscription-based business models for electronics, where buyers pay for access rather than ownership. For such platforms, accurate multi-horizon price forecasts are directly tied to financial viability: systematic overestimation exposes the business to losses when items depreciate faster than forecast, while systematic underestimation transfers risk to the customer and reduces service attractiveness.

\subsection*{Research gap}

Prior forecasting research has concentrated on structured markets - stocks, foreign exchange, electricity, and real estate - where data is abundant and exhibits clear seasonal or autoregressive patterns \parencite{aggarwalElectricityPriceForecasting2009, lagoForecastingDayaheadElectricity2021, mohdOverviewRealEstate2020}. For used goods, existing studies treat pricing as a static tabular regression task, focusing on cars \parencite{hankarUsedCarPrice2022, ganeshUsedCarsPrice2019, dutulescuCarPriceQuotes2023} and real estate \parencite{truongHousingPricePrediction2020, wangDeepLearningModel2021} without a temporal dimension. The few time-series studies that do exist apply only ARIMA on stable, high-volume Amazon product categories \parencite{cartaForecastingECommerceProducts2019, fathallaNovelPricePrediction2023} or rely on rich automotive covariate sets \parencite{sunResearchResidualValue2022} - settings that differ fundamentally from the sparse, volatile listing data of regional second-hand marketplaces \parencite{rouxSecondhandMarketsAlternative2020, hewamalageForecastEvaluationData2023b}.

This review reveals four specific, unaddressed gaps in the literature:
\begin{enumerate}
    \item \textbf{No time-series benchmark for used electronics.} No study has compared statistical and deep learning forecasting models on a real-world used electronics price dataset under controlled, reproducible conditions.
    \item \textbf{No evaluation of modern deep learning architectures in this domain.} Recent advances - including N-BEATS \parencite{oreshkinNBEATSNeuralBasis2020a}, N-HiTS \parencite{challuNHITSNeuralHierarchical2023}, and transformer-based models \parencite{limTemporalFusionTransformers2021, nieTimeSeriesWorth2023, zhouInformerEfficientTransformer2021c} - have not been evaluated on used goods data.
    \item \textbf{No multi-horizon evaluation.} Subscription pricing requires accurate forecasts from 1 to 365 days ahead, yet no existing study evaluates a model family across this full horizon range in any used goods domain.
    \item \textbf{No hyperparameter robustness or deployment analysis.} For automated pricing systems, understanding the sensitivity of model performance to configuration choices is critical for safe production deployment, yet it is unexplored in this context.
\end{enumerate}

\subsection*{Research hypotheses}

Based on the domain characteristics identified above, we formulate three testable hypotheses:

\textbf{H1 (Deep learning advantage).} When evaluated under consistent experimental conditions, deep learning models yield significantly lower forecasting errors than classical statistical benchmarks (ARIMA, ETS, Theta) across both short- and long-term horizons \parencite{baoDeepLearningFramework2017a, oreshkinNBEATSNeuralBasis2020a}. We test this by comparing Mean Absolute Percentage Error (MAPE) across model tiers at each horizon with pairwise statistical significance tests.

\textbf{H2 (Horizon degradation of advantage).} The advantage of deep learning over statistical models diminishes at longer horizons, as error accumulation and compounding uncertainty reduce the relative gain of more expressive architectures \parencite{oreshkinNBEATSNeuralBasis2020a, salinasDeepARProbabilisticForecasting2020}. We test this by examining whether the MAPE gap between tiers narrows or reverses as the horizon increases from 1 to 365 days.

\textbf{H3 (Complexity–stability tradeoff).} Deep learning models deliver superior average-case accuracy, but their sensitivity to hyperparameter choices imposes higher tuning effort and deployment risk compared to statistical methods \parencite{dewittOpenSourceReproducible2024b, hewamalageForecastEvaluationData2023b}. We test this through variance analysis of hyperparameter sensitivity and best-versus-mean MAPE gaps across tuning configurations.

\subsection*{Contributions}

This study addresses the four identified gaps through the following original contributions:

\begin{itemize}
    \item \textbf{First real-world dataset for used electronics time-series forecasting.} A large-scale proprietary collection of daily price listings for over 100 used smartphone and laptop models scraped from Polish online marketplaces (January 2022 to March 2025), preprocessed via fuzzy matching, Natural Language Processing (NLP)-based entity recognition, and time-series smoothing to form structured \textit{abstract-item} price series. This is the first dataset of this type designed specifically for multi-horizon time-series forecasting in the used electronics domain.
    \item \textbf{First systematic multi-horizon benchmark for this domain.} Eleven forecasting models - classical statistical (ARIMA, ETS, Theta, Na\"{i}ve), recurrent and convolutional (Long Short-Term Memory - LSTM, Temporal Convolutional Network - TCN), and modern deep architectures (N-BEATS, N-HiTS, Temporal Fusion Transformer - TFT, PatchTST, Informer) - are evaluated head-to-head across six horizons (1, 7, 30, 90, 180, 365 days) under identical experimental conditions, providing the first direct comparison in this market.
    \item \textbf{Novel three-protocol evaluation framework.} Three complementary protocols distinguish trajectory fitness (Protocol 1), one-shot terminal accuracy (Protocol 2), and cross-horizon model reuse (Protocol 3), enabling conclusions about whether a single far-horizon model can replace a family of horizon-specific production models - a question not addressed in prior benchmarking studies.
    \item \textbf{First hyperparameter robustness analysis for production pricing.} Variance analysis across tuning configurations quantifies sensitivity per model and horizon, providing actionable guidance on deployment risk for automated pricing systems - a criterion absent from existing forecasting benchmarks.
\end{itemize}

\section{Related work}
\label{sec:literature}

Researchers have devoted a lot of attention to predicting goods prices using various forecasting methods. 
We can observe statistical models predicting prices of cars, real estate, stocks, indexes, electricity and many others \parencite{aggarwalElectricityPriceForecasting2009, geertsSurveyMethodsInput2023, hankarUsedCarPrice2022a, lagoForecastingDayaheadElectricity2021, mohdOverviewRealEstate2020, samruddhiUsedCarPrice2020}.
Despite many valued commodities, little attention has been paid to second-hand electronics and other e-commerce items.

\subsection{Pricing data}

Most literature regarding used goods prediction revolves around used cars or houses.
Due to the abundance and openness of the vehicle ad lists, the data mainly consists of listings shared over the Internet, scrapped or collected manually, each with its own set of attributes and features, including i.a. mileage, fuel type, location, area or production year \parencite{ganeshUsedCarsPrice2019, bergmannMachineLearningPredicting2025, hankarUsedCarPrice2022, truongHousingPricePrediction2020}.
Furthermore, advancements in natural language processing (NLP) or image recognition, supported by deep learning, make feature engineering much more possible, enriching the feature space for predictive modelling by using i.a., satellite images or simply car photos \parencite{yangAIBlueBook2018, wangDeepLearningModel2021, dutulescuCarPriceQuotes2023}.

Despite abundant research in the automotive and household sectors, similar data usage for other categories of used goods, including electronics, remains relatively underdeveloped \parencite{fengDevelopmentChallengesSecondHand2024b}. 
Unlike the areas mentioned above, we lack comprehensive electronics datasets that could be available at a glance. 
While for households or cars, it is easy to identify an abstract item in an online listing, the diversity of the product brands and types, including different naming techniques between online marketplaces, makes it very difficult to correctly assign an item to the listing.
Most of the researchers refer to online platforms such as eBay \parencite{ebay} or OLX \parencite{olx}, creating crawlers that scrape data from selected categories, e.g. \cite{ghaniPricePredictionInsurance2005} scraped data about laptops and palmtops, \cite{raykhelRealtimeAutomaticPrice2009} scraped data about laptops, describing every listing with characteristics such as CPU, RAM or operating system.
\cite{fathallaDeepEndtoendLearning2020} have also acquired such a dataset, stating that they obtained 107,000 observations with descriptions, images and prices of the items from 24 different categories.
Some such datasets have been published on the Kaggle platform \parencite{MercariPriceSuggestion, AmazonSalesDataset}. 
Unlike the automotive and household markets, where centralized platforms aggregate listings from various sources, publishing data publicly, the electronics resale landscape relies on multiple regional platforms, including Facebook groups, each catering to specific geographic or demographic segments. 
As such, any attempt to develop machine learning models will fail without first obtaining data from these local sources.
However, data is usually vast and sparse since there are many categories of items, different properties of the goods, etc. 

\subsection{Price forecasting}

Most of the applications of used goods price forecasting revolve around tabular data.
For the automotive market, researchers have applied mostly all new machine learning algorithms, including linear regression, random forest, gradient boosting or neural networks \parencite{hankarUsedCarPrice2022, ganeshUsedCarsPrice2019, dutulescuCarPriceQuotes2023, palHowMuchMy2017} obtaining excellent results measured in $R^2$ or mean average percentage error (MAPE).
The same conclusions can be made for the household market \parencite{wangDeepLearningModel2021, truongHousingPricePrediction2020}.
Regarding used electronics, we can refer to \cite{aliExtensiveDataExploration2018}, who solved one of the Kaggle problems.
In their solution, they created a tabular set on top of which they used a neural network, obtaining a root mean squared log error of 0.5001.
Other approaches include k-nearest neighbours proposed by \cite{raykhelRealtimeAutomaticPrice2009} or deep learning by \cite{zecResidualValuePrediction2022}.
Most of these experiments were conducted either on open-source automotive datasets or on independently scraped datasets.

However, the problem is also tackled from other types of data sources.
Many price prediction models employ NLP techniques to extract features from textual descriptions of the used goods.
Most approaches utilize sentiment estimation that is later correlated with the price to enhance the model \parencite{tsengPricePredictionEcommerce2018, nicholsonNovelMethodPredicting2013}.
Since the used goods market is related to the stock market, we can also relate to studies in this domain.
Researchers utilize topic modelling or deep learning architectures based on embeddings, offering insights into market dynamics. 
Existing research \parencite{andrawosNLPStockMarket2022, maneStockMarketPrediction2022, puhPredictingStockMarket2023, ashtianiNewsbasedIntelligentPrediction2023} highlights the importance of incorporating textual data from various sources, including financial news articles, social media posts, and analyst reports.
On the other hand, academics also try to find specific parameters of the items in the textual corpora (descriptions, opinions), e.g. finding word stems such us 'break' or 'damage', could influence price significantly \parencite{pudaruthPredictingPriceUsed}.
In another paper \cite{fuchsProductTitlestoAttributesTexttoText2022}, the authors bring up the subject of identifying a set of attribute-value pairs for products on e-commerce services, which would be the most descriptive with a Text-to-Text method that focuses on tracking users' search history which prioritizes listing and filtering on e-Bay to retrieve a set of attribute-value pairs for products.
Not only text data has been extensively used to predict used goods prices.
\cite{fathallaDeepEndtoendLearning2020} and \cite{hanPriceSuggestionOnline2020} propose a multimodal approach to forecasting prices, including both textual and visual features.
The use of image recognition in this field is much more often seen for used car prediction problems, where the state of the car dramatically impacts the price estimate.
\cite{ahmedHousePriceEstimation2016} analyzed how the inclusion of visual features of a household can affect the accuracy of house price forecasts, revealing a significant correlation between forecast accuracy and the number of visual features included.

In cases where the price prediction does not relate to one entity but rather to a group of products, e.g. general price of used iPhones 16 over time, a time series prediction is usually proposed.
First and foremost, product prices change over time -- focusing only on the listings can provide inaccurate predictions at the moment of forecasting due to the use of too old quotations.
Models that do not use time factor as one of their features should not be considered accurate \parencite{fathallaDeepEndtoendLearning2020}.
Once again, more research has been done for the automotive and household markets, where we can observe approaches employing machine learning time series data models predicting second-hand car prices \parencite{sunResearchResidualValue2022} or predictions of the price index of second-hand houses with neural networks \parencite{xuSecondhandHousePrice2022, maResearchApplicationSecondHand2021}.
Regarding used electronics, we can mention standard approaches utilized by \cite{cartaForecastingECommerceProducts2019} and \cite{fathallaNovelPricePrediction2023}, who both leverage ARIMA models predicting time series values for different Amazon products, e.g. books, electronics or movies and TV.
In the area of e-commerce, \cite{osawaProductSalesForecast2021} has proposed an ARIMA model trained on the customers' purchase history including seasonal fluctuations in product prices. In contrast, \cite{smedmarkTrendDetectionTime2021} suggests neural networks tackle the trend analysis.
We can once again refer to stock price prediction, which appears similar to used electronics pricing in this context due to their shared volatility and sensitivity to market dynamics.
One of the most comprehensive financial time series forecasting reviews has been provided by \cite{sezerFinancialTimeSeries2019}.
They provide a range of deep learning models that include both univariate and multivariate approaches, both with the datasets used and the results obtained.
Among the most robust approaches, we can mention Neural Hierarchical Interpolation for Time Series (N-HiTS) \parencite{challuNHITSNeuralHierarchical2023} and Neural Basis Expansion Analysis for Time Series (N-BEATS) \parencite{oreshkinNBEATSNeuralBasis2020a}.

More recently, the research community has shifted toward time series foundation models - large pre-trained transformers capable of zero-shot or few-shot forecasting across diverse domains. Chronos \parencite{ansariChronosLearningLanguage2024} tokenizes time series and fine-tunes pre-trained language model architectures, achieving competitive zero-shot performance on 42 public benchmarks. TimesFM \parencite{dasDecoderOnlyFoundationModel2024} is a decoder-only model pre-trained on a large corpus of real and synthetic time series, approaching supervised accuracy without task-specific training. MOIRAI \parencite{wooUnifiedTrainingInference2024} unifies training across heterogeneous time series datasets via a multi-patch tokenization scheme. The iTransformer \parencite{liuiTransformerEffectiveTransformer2024} re-examines the role of attention in transformers by applying it across variates rather than time steps, improving multivariate forecasting on standard benchmarks. These developments underscore the rapidly growing capability of deep learning for time series, yet - as our study demonstrates - their advantage is not universal and depends critically on the nature of the target domain.


Table~\ref{tab:litreview} summarises the most relevant studies surveyed in this section, highlighting the domain, approach, forecasting task type, and key limitation with respect to the present work.

\begin{table}[htbp]
\centering
\footnotesize
\caption{Summary of surveyed studies on used goods price prediction and related time-series forecasting}
\label{tab:litreview}
\begin{tabularx}{\textwidth}{>{\raggedright\arraybackslash}p{3.2cm}|>{\raggedright\arraybackslash}p{2.2cm}|>{\raggedright\arraybackslash}p{2.0cm}|>{\raggedright\arraybackslash}p{2.0cm}|>{\raggedright\arraybackslash}X}
\toprule
\textbf{Reference} & \textbf{Domain} & \textbf{Approach} & \textbf{Task} & \textbf{Gap / Limitation} \\
\midrule
\cite{hankarUsedCarPrice2022}, \cite{ganeshUsedCarsPrice2019}, \cite{dutulescuCarPriceQuotes2023} & Used cars & ML (RF, NN, XGBoost) & Tabular regression & No time dimension; point-in-time price only \\
\cite{truongHousingPricePrediction2020}, \cite{wangDeepLearningModel2021} & Real estate & ML / DL + images & Tabular regression & Structured market, rich listing attributes \\
\cite{ghaniPricePredictionInsurance2005}, \cite{raykhelRealtimeAutomaticPrice2009} & Used electronics (laptops) & Statistical, KNN & Tabular regression & Small scale; no time-series evaluation \\
\cite{aliExtensiveDataExploration2018} & Used electronics (Kaggle) & NN & Tabular regression & Single competition dataset; no temporal modelling \\
\cite{fathallaDeepEndtoendLearning2020} & Multi-category e-commerce & Multimodal DL & Tabular regression & Features per listing; no aggregate time-series \\
\cite{cartaForecastingECommerceProducts2019}, \cite{fathallaNovelPricePrediction2023} & Amazon products & ARIMA & Time-series & Single model; no DL comparison; stable market \\
\cite{osawaProductSalesForecast2021} & E-commerce sales & ARIMA & Time-series & Sales volume, not price; seasonal and structured \\
\cite{sunResearchResidualValue2022} & Used car residual value & ML time-series & Time-series & Automotive; rich covariate set \\
\cite{sezerFinancialTimeSeries2019} & Financial markets & DL survey & Time-series & Structured, high-liquidity markets only \\
\cite{oreshkinNBEATSNeuralBasis2020a}, \cite{challuNHITSNeuralHierarchical2023} & General time-series & N-BEATS, N-HiTS & Multi-horizon & Not evaluated on sparse, non-standard goods data \\
\bottomrule
\end{tabularx}
\end{table}

As the table illustrates, prior work either treats price prediction as a static tabular problem without a time dimension, or applies time-series models to structured markets with abundant and regular data. Studies specifically addressing used electronics as a time-series forecasting problem are limited to ARIMA applied to stable, high-volume Amazon product categories - settings that differ fundamentally from the sparse, volatile listing data of second-hand marketplaces. Furthermore, recent advances in deep learning for time series - including the N-BEATS and N-HiTS architectures \parencite{oreshkinNBEATSNeuralBasis2020a, challuNHITSNeuralHierarchical2023}, transformer-based models such as PatchTST \parencite{nieTimeSeriesWorth2023} and TFT \parencite{limTemporalFusionTransformers2021}, and emerging time-series foundation models \parencite{salinasDeepARProbabilisticForecasting2020} - have not been systematically evaluated in this domain. The present study fills this gap by providing the first multi-horizon benchmark of both classical and modern deep learning architectures on a real-world used electronics dataset.

\section{Data Collection and Preprocessing}
\label{sec:data}

Our dataset consists of scraped daily listings from social media and publicly available offer sites in Poland that facilitate the sale of used goods. Since these platforms do not provide APIs, we employ web scraping techniques to capture the broadest possible market representation. The dataset is constructed to enable the forecasting of abstract item prices by aggregating price listings over time, where items are identified without specific individual characteristics. An \textit{abstract item} refers to a generalized product representation, such as “Apple iPhone 14 Pro 256GB”, that groups together multiple listings of the same model rather than tracking individual physical units.

To ensure accurate price tracking, we developed a system that matches abstract item names (e.g., Apple iPhone 11 Pro Max 512GB) to specific listings that may not be directly comparable. Sellers often use varied titles - such as \textit{iPhone 11 ProMax 512 GB}, \textit{Apple 11Pro Max 512G}, or \textit{iPhone 11PM 512GB graphite} - which differ in abbreviations, spacing, and word order but refer to the same product. Matching is performed using a combination of keyword-based scoring, fuzzy string matching, Named Entity Recognition (NER) models, and text embeddings. Each listing is assigned a relevance score, and a predefined confidence threshold filters out low-quality matches. Sanity checks prevent systematic errors such as merging \textit{iPhone Pro} and \textit{iPhone Pro Max} listings into a single category. The complete matching pipeline is described in a separate study; the present work uses its output as input to the forecasting framework.

\begin{figure}
    \centering
    \includegraphics[width=\linewidth]{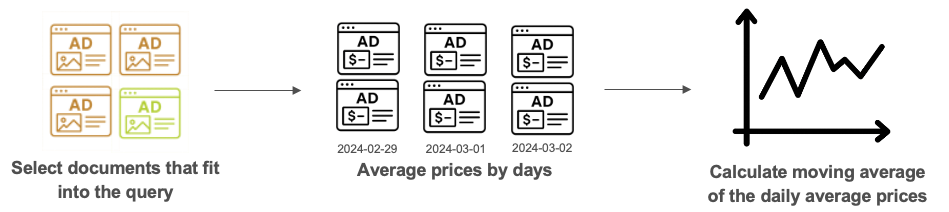}
    \caption{A schema of the time series preparation system}
    \label{fig:graphic-depiction}
\end{figure}

Once matched, daily price listings for each abstract item are aggregated using a simple mean, forming a structured time series. To mitigate distortions caused by extreme values, we exclude outliers by removing listings priced above 30{,}000~PLN or below 300~PLN (thresholds set to match the subscription business’s operational price range), as well as observations exceeding $\pm3$ standard deviations from the 30-day rolling mean. A 30-day trailing moving average is then applied to smooth daily volatility arising from irregular listing activity while preserving longer-term price trends; a shorter window would retain too much noise, while a longer window would lag behind genuine structural price changes. Missing values - arising when no listings exist for a given item on a given day - are imputed by forward-filling from the most recent available observation. Forward-filling is conservative in the sense that it does not introduce synthetic variation, but it does attenuate true volatility in sparse series. This effect is discussed further in Section~\ref{sec:summary}. A graphical depiction of this matching and aggregation process is presented in Figure~\ref{fig:graphic-depiction}.

For this study, we focus on smartphones and laptops, although the system is designed to generalize to other product categories. The specific abstract items selected for forecasting reflect a bias toward high-end and Apple-branded devices, aligning with the company’s core sales inventory. Readers should note that this selection may inflate the apparent performance of all models - high-value, frequently traded items tend to exhibit smoother and more predictable price trajectories than low-end or niche devices with sparse listing histories. Model rankings may therefore differ on datasets with greater category diversity or lower average price levels. All models are tested within the same dataset constraints, ensuring that reported comparisons remain internally valid despite this sampling bias.

To validate the effectiveness of the system, we compare automated pricing results -- calculated as the last observation from the time series -- to actual business prices of sold used goods. The details of the matching algorithm and its full implementation are covered in a separate study. Table~\ref{table:pricing-results} presents exemplary results of the pricing process, showing the best predicted value per item generated by the matching algorithm alongside the corresponding business (market) price.

\begin{table}[h]
\centering
\caption{Exemplary results of pricing using our matching algorithm with the best prediction per item compared with its business value}
\label{table:pricing-results}

\begin{tabularx}{0.75\textwidth}{>{\raggedright\arraybackslash}Xcc}
\toprule
\textbf{Abstract item} & \textbf{Predicted price} & \textbf{Market price} \\
\midrule
apple iphone 13 pro max 512gb grafitowy & 4306.70 & 4250 \\[1ex]
samsung galaxy a13 128gb               & 730.45  & 400  \\[1ex]
samsung galaxy s22 ultra 256gb         & 3288.09 & 3200 \\[1ex]
acer aspire a515 4gb 128gb             & 1203.23 & 1150 \\[1ex]
apple macbook pro m2 512gb space gray  & 5416.09 & 6000 \\[1ex]
\bottomrule
\end{tabularx}

\end{table}

For this study we prepared a dataset containing time series of prices from 01-01-2022 to 31-03-2025 for a set of most popular items from two categories: laptops and smartphones, including models such as Apple MacBooks, ASUS, Acer and HP laptops, and Samsung and Apple smartphones that were actively traded on Polish used goods markets during that period. Table~\ref{tab:dataset_stats} summarises the key quantitative characteristics of the dataset.

\begin{table}[H]
\centering
\caption{Quantitative characteristics of the dataset used in this study}
\label{tab:dataset_stats}
\begin{tabularx}{0.85\textwidth}{>{\raggedright\arraybackslash}X r}
\toprule
\textbf{Statistic} & \textbf{Value} \\
\midrule
Date range & January 2022 -- March 2025 \\
Number of calendar days (series length) & 1,186 \\
Unique abstract items & 357 \\
\quad of which smartphones & 229 \\
\quad of which laptops & 127 \\
\quad of which other & 1 \\
Total item-day observations & 376,754 \\
Total raw listings aggregated & $\approx$11.8 million \\
\midrule
\textbf{Observations per series} & \\
\quad Minimum & 36 \\
\quad Median & 1,186 \\
\quad Mean & 1,055 \\
\quad Maximum & 1,186 \\
\midrule
\textbf{Missing days per series (sparsity, before imputation)} & \\
\quad Mean & 3.9\% \\
\quad Median & 0.1\% \\
\quad Maximum (most sparse series) & 75.4\% \\
\quad Overall fraction missing & 3.7\% \\
\midrule
\textbf{Price distribution (PLN, post-filter, pre-smoothing)} & \\
\quad Mean & 3,258 \\
\quad Median & 2,600 \\
\quad Standard deviation & 2,453 \\
\quad 10th percentile & 1,115 \\
\quad 90th percentile & 6,012 \\
\quad Raw listings per item-day (median) & 11 \\
\bottomrule
\end{tabularx}
\end{table}

\begin{figure}[H]
    \centering
    \includegraphics[width=\linewidth]{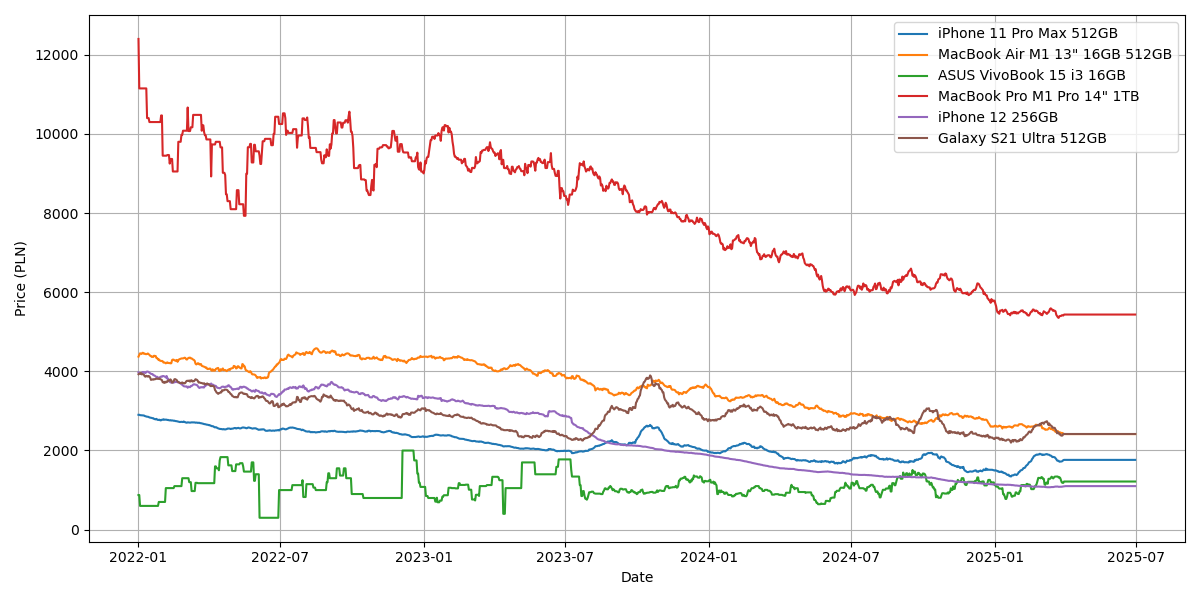}
    \caption{Exemplary historical postprocessed time series of used goods analysed in this paper}
    \label{fig:fig1}
\end{figure}

The dataset exhibits highly heterogeneous sparsity: while the median series has virtually no missing days (0.1\%), a small number of niche items have up to 75.4\% of days without any listing. The price distribution is right-skewed, reflecting the dominance of high-end devices (mean 3,258~PLN vs.\ median 2,600~PLN). This combination of sparse, volatile, right-skewed series distinguishes the dataset from the structured financial benchmarks on which most deep learning time-series architectures were originally evaluated. Figure~\ref{fig:fig1} presents exemplary processed time series for selected items.

\section{Methodology}
\label{sec:methods}

\subsection{Models}

Model selection follows a generational structure designed to provide interpretable comparisons across the spectrum of forecasting complexity. The first tier comprises classical statistical methods (ARIMA, ETS, Theta, Na\"{i}ve), which serve as well-established benchmarks with minimal hyperparameter requirements and known theoretical properties. The second tier includes recurrent and convolutional neural networks (LSTM, TCN), representing the first wave of deep learning applied to time series and commonly used in the used-goods and financial forecasting literature \parencite{sezerFinancialTimeSeries2019}. The third tier covers modern deep architectures: decompositional models (N-BEATS, N-HiTS), which have achieved state-of-the-art results in general time-series competitions \parencite{oreshkinNBEATSNeuralBasis2020a, challuNHITSNeuralHierarchical2023}, and transformer-based models (TFT, PatchTST, Informer), which have been widely adopted in long-horizon forecasting research \parencite{limTemporalFusionTransformers2021, nieTimeSeriesWorth2023, zhouInformerEfficientTransformer2021c}.

Each model receives as input a univariate historical time series representing the daily smoothed price of a specific electronic item. Statistical baselines (ARIMA, ETS, Theta, Na\"{i}ve) generate multi-step forecasts recursively - producing one step at a time and feeding each prediction back as input. All deep learning models (LSTM, TCN, N-BEATS, N-HiTS, TFT, PatchTST, Informer) are configured as direct multi-output models: they predict all $h$ future time steps simultaneously in a single forward pass, which avoids compounding recursive errors at the cost of requiring horizon-specific training.

\subsubsection{Long Short-Term Memory (LSTM)}

The Long Short-Term Memory (LSTM) network \parencite{hochreiterLongShortTermMemory1997b} is a recurrent neural network (RNN) variant designed to capture long-range temporal dependencies by gating information flow through forget, input, and output gates. This mechanism mitigates the vanishing gradient problem that limits standard RNNs. In our setup, a stacked LSTM encoder processes the historical price window $\{y_{t-p}, \dots, y_{t-1}\}$ and a fully connected decoder maps the final hidden state to the $h$-step forecast vector, trained end-to-end with mean squared error loss.

\subsubsection{Neural Basis Expansion Analysis for Time Series (N-BEATS)}
The N-BEATS model \parencite{oreshkinNBEATSNeuralBasis2020c} follows a backcast-forecast decomposition framework, where the model simultaneously reconstructs past values and predicts future ones. This decomposition enables N-BEATS to effectively model trends, seasonality, and noise without requiring prior knowledge of the time series structure.

N-BEATS is composed of multiple stacked blocks, each containing a fully connected four-layer feedforward network with Rectified Linear Unit (ReLU) activation functions, denoted as $\textbf{FCNN}_k$. Given an input time series segment $\mathbf{x}_k$, each block generates expansion coefficients $\theta_k^b$ and $\theta_k^f$, which are then projected onto basis functions $g_k^b(\cdot)$ and $g_k^f(\cdot)$ to produce backcast and forecast components, as in equation~\ref{eq:eq1}.

\begin{equation}
\begin{aligned}
\label{eq:eq1}
h_k &= FCNN_k(x_k), \\
\theta_k^b &= linear^b_k(h_k), \quad \theta_k^f = linear^f_k(h_k), \\
\hat{\mathbf{x}}_k &= V_k^b \theta_k^b, \quad \hat{\mathbf{y}}_k = V_k^f \theta_k^f.
\end{aligned}
\end{equation}

Here, $linear$ denotes a linear layer, while $V_k^b$ and $V_k^f$ represent basis vectors used to reconstruct past values ($\hat{\mathbf{x}}_k$) and forecast future values ($\hat{\mathbf{y}}_k$). For an identity basis function, the forecast basis vector simplifies to $V_k^f = \mathds{1}_{p \times p}$.

N-BEATS employs a residual stacking mechanism to refine its predictions across multiple blocks. Each block generates a backcast estimate $\hat{\mathbf{x}}_k$, which is subtracted from the input time series $\mathbf{x}_k$ to create a residual $\mathbf{x}_{k+1}$:

\begin{equation}
\mathbf{x}_{k+1} = \mathbf{x}_k - \hat{\mathbf{x}}_k.
\end{equation}

This residual is then passed to the next block, allowing subsequent layers to focus on components of the time series that were not fully captured. The final forecast is computed as the sum of outputs from all blocks:

\begin{equation}
\hat{\mathbf{y}} = \sum_{k=1}^{K} \hat{\mathbf{y}}_k,
\end{equation}

where $K$ is the total number of stacked blocks in the architecture.

\subsubsection{Neural Hierarchical Interpolation for Time Series (N-HiTS)}
The N-HiTS \parencite{challuNHiTSNeuralHierarchical2022a} model builds upon the block-based structure of N-BEATS but introduces hierarchical interpolation mechanisms to enhance predictions across multiple resolutions. The key innovations of N-HiTS lie in its multi-rate signal sampling and temporal interpolation, allowing it to capture time series patterns at different scales.

Unlike N-BEATS, where each block processes the raw time series input $\mathbf{x}_k$ through a feedforward network, N-HiTS first downsamples the input using a MaxPool layer with multiple kernel sizes $ks_k$. This approach, referred to as multi-rate signal sampling, enables different blocks to operate at different temporal resolutions, effectively learning various aspects of the time series at different scales, as in equation~\ref{eq:eq2}.

\begin{equation}
\begin{aligned}
\label{eq:eq2}
x_k^{(p)} &= MaxPool(x_k, ks_k), \\
h_k &= FCNN_k(x_k^{(p)}), \\
\theta_k^b &= linear^b_k(h_k), \quad \theta_k^f = linear^f_k(h_k).
\end{aligned}
\end{equation}

This mechanism allows each block to extract information relevant to its respective time scale, improving the model’s ability to capture both short- and long-term dependencies.

N-HiTS also introduces temporal interpolation to manage the dimensionality of expansion coefficients, which depend on the input length. To address this, the model defines an expressiveness ratio $r_k$, which controls how many additional time units need to be interpolated. The forecast and backcast components are then generated using an interpolation function $g(\cdot)$ that maps the expansion coefficients $\theta_k^b$ and $\theta_k^f$ to the time domain, as in equation~\ref{eq:eq3}.

\begin{equation}
\begin{aligned}
\label{eq:eq3}
\hat{\mathbf{x}}_k &= g(\tau, \theta_k^b), \
\hat{\mathbf{y}}_k &= g(\tau, \theta_k^f).
\end{aligned}
\end{equation}

Here, $\tau$ represents the time index, and the hierarchical nature of this interpolation ensures that the expressiveness ratio $r_k$ for each block is synchronized with the kernel size $ks_k$ used in the pooling layer.

As in N-BEATS, the final forecast is obtained by summing the forecasts from all stacked blocks, as in equation~\ref{eq:eq4}.

\begin{equation}
\label{eq:eq4}
\hat{\mathbf{y}} = \sum_{k=1}^{K} \hat{\mathbf{y}}_k.
\end{equation}

\subsubsection{Temporal Convolutional Network (TCN)}

The Temporal Convolutional Network (TCN) \parencite{baiEmpiricalEvaluationGeneric2018b} replaces recurrence with stacked dilated causal convolutions, achieving an exponentially growing receptive field without sequential computation. A dilated convolution at layer $l$ with dilation factor $d$ and kernel size $K$ is defined as $h_t^{(l)} = \sigma\!\left(\sum_{k=0}^{K-1} W_k^{(l)} \cdot x_{t-k \cdot d} + b^{(l)}\right)$, where causality is enforced by left-sided padding. A fully connected decoder maps the final representation to the $h$-step forecast vector.

\subsubsection{Temporal Fusion Transformer (TFT)}

TFT \parencite{limTemporalFusionTransformers2021} combines variable selection networks, an LSTM encoder-decoder for short-term pattern extraction, and multi-head self-attention for long-range dependencies. Gated Residual Networks (GRNs) control information flow throughout the architecture, providing interpretable feature importance. In our univariate setting, static covariates are absent; TFT is trained with a pinball loss at the median quantile ($\tau = 0.5$), yielding point forecasts.

\subsubsection{PatchTST}

PatchTST \parencite{nieTimeSeriesWorth2023} segments the input series into fixed-length overlapping patches of size $P$ with stride $S$, projecting each patch $\mathbf{p}_i \in \mathbb{R}^P$ to a $d$-dimensional token via $\mathbf{z}_i = \mathbf{W}_e \mathbf{p}_i + \mathbf{b}_e$. Positional encodings are added and the resulting sequence is processed by a standard transformer encoder with multi-head self-attention. A linear prediction head maps the final token representations to the $h$-step forecast, with patch tokenization reducing sequence length while preserving local temporal structure.

\subsubsection{Informer}

Informer \parencite{zhouInformerEfficientTransformer2021c} addresses the quadratic self-attention bottleneck through ProbSparse attention, which retains only the $u$ queries with highest Kullback-Leibler (KL) divergence from uniform attention, reducing complexity to $\mathcal{O}(L \log L)$. A self-distillation step halves the sequence length after each encoder layer via convolution and max-pooling: $\mathbf{H}^{(l+1)} = \text{Pool}(\text{Conv}(\mathbf{H}^{(l)}))$. A generative-style decoder produces all forecast steps in a single forward pass.

\subsubsection{Statistical baselines}

We include four classical baselines widely used in forecasting competitions \parencite{makridakisM3CompetitionResultsConclusions2000a, makridakisM5AccuracyCompetition2022} for their simplicity and strong empirical performance.

\textbf{Na\"{i}ve.} The last observed value is propagated forward: $\hat{y}_{t+h} = y_t$ for $h = 1, \dots, H$.

\textbf{ARIMA$(p,d,q)$} \parencite{boxJenkinsReinselLjung2015} combines autoregressive ($p$), integration ($d$), and moving average ($q$) components. After $d$-order differencing $y_t’ = (1-B)^d y_t$, the model is:
\begin{equation}
    y_t’ = \sum_{i=1}^p \phi_i y_{t-i}’ + \varepsilon_t + \sum_{j=1}^q \theta_j \varepsilon_{t-j},
\end{equation}
where $\phi_i$ are autoregressive coefficients, $\theta_j$ are moving average coefficients, $B$ is the backshift operator ($B^k y_t = y_{t-k}$), and $\varepsilon_t$ is white noise. Orders are selected per series by minimising the Akaike Information Criterion (AIC) over a search grid.

\textbf{ETS} \parencite{hyndmanForecastingExponentialSmoothing2008} models the error (E), trend (T), and seasonal (S) components, each of which may be absent (N), additive (A), or multiplicative (M). The best configuration is selected automatically by AIC. For example, the additive ETS(A,A,A) model updates:
\begin{equation}
\begin{aligned}
\ell_t &= \ell_{t-1} + b_{t-1} + \alpha \varepsilon_t, \quad
b_t = b_{t-1} + \beta \varepsilon_t, \quad
s_t = s_{t-m} + \gamma \varepsilon_t,
\end{aligned}
\end{equation}
where $\alpha, \beta, \gamma \in [0,1]$ are smoothing parameters, $\ell_t$ is the level, $b_t$ is the trend, $s_t$ is the seasonal component with period $m$, and $\varepsilon_t$ is white noise. $h$-step forecasts follow $\hat{y}_{t+h} = \ell_t + hb_t + s_{t+h-m(\lfloor(h-1)/m\rfloor+1)}$.

\textbf{Theta} \parencite{astanakopoulosTheta2000} decomposes the series into two modified theta lines obtained by applying second differences with coefficients $\theta=0$ and $\theta=2$ to control curvature. The $\theta=0$ line captures the long-term linear trend via ordinary least squares; the $\theta=2$ line amplifies short-term dynamics and is extrapolated with Simple Exponential Smoothing (SES). The final forecast is the average of both components, a strategy that won the M3 forecasting competition \parencite{makridakisM3CompetitionResultsConclusions2000a}.

\subsection{Performance criteria}\label{sec:cross-valid}
Our predictive framework features meticulously specified criteria for model evaluation, taking into account both business and engineering requirements. This methodology is employed in both the experimental and production environments to select the best models and fine-tune their hyperparameters. Unlike standard approaches that typically rely solely on error-based metrics such as MAPE or Mean Squared Error (MSE), our framework provides a broader performance assessment, incorporating additional statistical and operational indicators. Moreover, during cross-validation, we apply a champion–challenger strategy, where the current production model (champion) is continuously compared against newly trained candidates (challengers) to ensure consistent improvement and stability in real-world deployment.

\subsubsection{Evaluation and tuning strategy}
Our fundamental assumption during the design of the evaluation strategy is to replicate the requirements of a real-world business environment. Within this framework, predictive models with a specified time horizon for forecasts are recalculated at a frequency that prevents them from degrading due to model or data drift. Therefore, the primary evaluation technique we employ is Temporal Double Cross Validation based on the Forward Chaining approach, see Algorithm \ref{algo:temporal-cv} (please note that the data is stored in wide format, where each column represents a separate item). This algorithm allows us to evaluate our models $M$ within a specified forecast horizon $h$ across $w$ out of sample (OOS) time windows with a step size of $s$ between each window. For instance, for a forecast with $h=365$ days, we use $w=3$ cross-validation windows with intervals between forecasts set at $s=30$ days. In addition to that, in each cross-validation OOS window, we run hypertuning, looking for a best set of hyperparameters, evaluating performance over set of OOS windows, without retraining and expanding training sample. 

\begin{algorithm}[H]
\scriptsize
\caption{Temporal Cross-Validation - Forward Chaining}
\label{algo:temporal-cv}
\begin{algorithmic}[1]
    \Require{Panel dataset $D$, Prediction model $M$, Forecast horizon $h$, Number of CV windows $w$, Number of OOS windows $o_w$, Step size between each CV window $s$}
    \Ensure{Forecasts per each window}
    \Function{TemporalCV}{$D, M, h, w, o_w, s$}
        \State $m \gets$ \text{Number of data points (rows) in $D$}
        \State $n \gets$ \text{Number of items (columns) in $D$}
        \State $train\_size \gets (m - h)-(w-1) \times s$

        \State $forecasts \gets \{\}$ 

        \For{$i \gets 1$ to $w$}
            \State $train\_data \gets D[0 : train\_size, n]$
            \State $test\_data \gets D[train\_size: train\_size + h, n]$
            \State $hyperparams \gets \{\}$
            \For{$o \gets 1$ to $o_w$}
                \State $train\_data \gets D[0 : train\_size - o_w, n]$
                \State $test\_data \gets D[train\_size: train\_size - o_w + o, n]$
                \State $hyperparams[o] \gets \Call{HyperTuneModel}{M, train\_data}$
            \EndFor
            \State $best\_hyperparams \gets \Call{Evaluate}{hyperparams}$
            \State \Call{TrainGlobalModel}{$M, best\_hyperparams, train\_data$}
            \State $forecast \gets \Call{Predict}{M, test\_data} $ \algorithmiccomment{forecast per each item}
            \State $forecasts[i] \gets forecast$

            \State $train\_size \gets train\_size + s$
            
        \EndFor

        \State \Return $forecasts$
    \EndFunction
\end{algorithmic}
\end{algorithm}

Application of Temporal Double Cross Validation on multiple models leads to obtaining forecasts with a very high granularity; that is, we achieve dimensions: $model \times window \times item$. To comprehensively understand the behaviour of the models, we need to reduce the dimensions of the analysis to $model \times window$. Thus, we apply straightforward scoring and aggregation with Algorithm \ref{algo:agg_scoring}. At crucial moments, such as choosing the best model architecture, we employ an additional aggregation level by averaging the evaluation metrics results at the $model$ level. This approach enables us to make well-defined decisions.
\begin{figure}[H]
    \centering
        \caption{Illustrative definition of temporal cross-validation utilized in this paper}
    \includegraphics[width=0.85\linewidth]{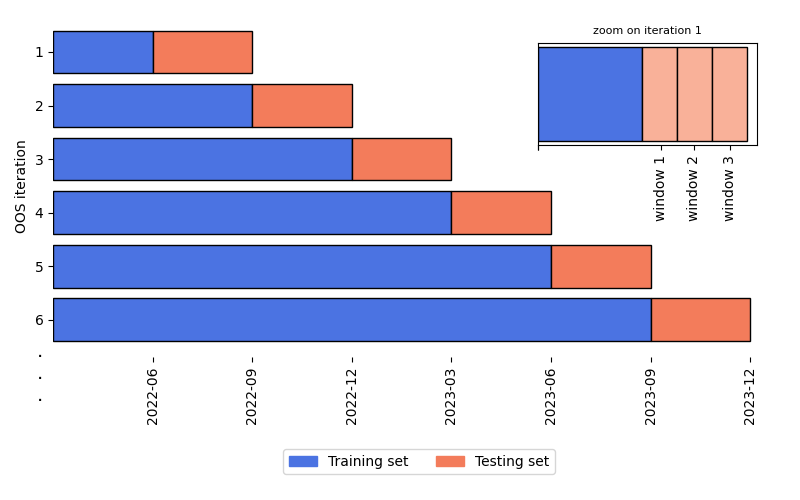}
    \label{fig:cross-validation}
\end{figure}

\begin{algorithm}[H]
\scriptsize
\caption{Aggregating scoring algorithm}
\label{algo:agg_scoring}
\begin{algorithmic}[1]
\Require{forecasts (per $model \times window \times item$), actuals (per $window \times item$), evaluation metrics $e$}
\Ensure{Aggregate score}
\Function{AggScore}{$\text{forecasts}$, $\text{actuals}$}
    \State $score \gets \{\} $
    \For{$\text{model} \in \text{models}$}
        \For{$\text{window} \in \text{windows}$}
            \For{$\text{item} \in \text{items}$}
                \State $f \gets \text{forecasts[model][window][item]}$
                \State $a \gets \text{actuals[window][item]}$
                \State $\text{score\_per\_item} \gets  \Call{ApplyEvaluationMetrics}{f, a, e}$
                \State $score[model][window][item] \gets \text{score\_per\_item}$
            \EndFor
            \State $\text{score\_per\_window} \gets \Call{Average}{score[model][window]}$
            \State $scores[model][window] \gets \text{score\_per\_window}$
        \EndFor
            \State $\text{score\_per\_model} \gets \Call{Average}{score[model]}$
            \State $scores[model] \gets \text{score\_per\_model}$
    \EndFor
        \State \textbf{return} $scores$
\EndFunction
\end{algorithmic}
\end{algorithm}


\begin{algorithm}[H]
\scriptsize
\caption{Best architecture searching and hyperparameters tuning}
\label{algo:tuning}
\begin{algorithmic}[1]
\Require{Set of prediction models architectures \{$M_1$, $M_2$, ..., $M_i$\}, Set of hyperparameter spaces per model architecture \{$H_1$, $H_2$, ..., $H_i$\}, Candidates searching algorithm $S$}
\Ensure{Best architecture and its hyperparameters}
\Function{ModelTuner}{$\text{forecasts}$, $\text{actuals}$}
    \State $scores \gets \{\} $
    \State $hps \gets \{\} $
    \For{$\text{model} \in$ \{$M_1$, $M_2$, ..., $M_i$\}}
        \State $\text{best\_hyperparameters, best\_score}  
        \gets  \Call{ApplyTunerHP}{ \newline
        \hspace*{5em} \text{hp\_space}=H_i,\newline
        \hspace*{5em} \text{cv\_strategy}=TemporalCV(\ldots),\newline
        \hspace*{5em} \text{score\_function} = AggScore(\ldots, e=MAPE), \newline
        \hspace*{5em} \text{hyperparams\_aggregation\_function} = AggHParams(\ldots, e=mean), \newline
        \hspace*{5em} \text{search\_algorithm}=S
        }$
        \State $hps[model] \gets best\_hyperparameters$
        \State $scores[model] \gets best\_score$
    \EndFor
    \State $champion\_model \gets model \text{ with minimal value in } scores$
    \State $champion\_hyperparameters \gets hps[champion\_model]$
        \State \textbf{return} $champion\_model$, $champion\_hyperparameters$
\EndFunction
\end{algorithmic}
\end{algorithm}

\subsubsection{Evaluation metrics}

Metric selection is driven by three business requirements. First, metrics must be interpretable to non-technical stakeholders who act on pricing recommendations: percentage-based errors are preferred over absolute ones because they are scale-independent across products with very different price levels (e.g., a 300~PLN smartphone vs.\ a 6000~PLN MacBook). Second, metrics must handle outliers explicitly: a single mismatched listing can produce extreme price observations, and a metric that is dominated by such events may not reflect typical business performance. Third, the metric suite must capture directional bias separately: overestimation causes financial losses when the subscription price is set too high and the item depreciates faster than forecast; underestimation transfers risk to the customer and reduces service attractiveness.
\begin{table}[H]
    \centering
    \caption{Evaluation metrics utilized in the framework}
    \label{tab:eval_metrics}
    \renewcommand{\arraystretch}{1} 
    \begin{tabularx}{\textwidth}{>{\raggedright\arraybackslash}X|>{\centering\arraybackslash}m{0.45\textwidth}}
    \toprule
    \textbf{Metric name} & \textbf{Formula} \\ 
    \midrule

    Mean Absolute Percentage Error (MAPE) & 
    \begin{equation}
        \frac{1}{n} \sum_{i=1}^{n} \left|\frac{F_i - A_i}{A_i}\right| = \frac{1}{n} \sum_{i=1}^{n} \text{APE}_i
    \end{equation} \\

    Winsorized MAPE (W-MAPE) & 
    \begin{equation}
        \frac{1}{n} \sum_{i=1}^{n} \text{winsorize}\!\left(\text{APE}, \text{limits}=[0.1, 0.1]\right)
    \end{equation} \\

    Symmetric Mean Absolute Percentage Error (SMAPE) &
    \begin{equation}
        \frac{1}{n} \sum_{i=1}^{n} \frac{2|F_i - A_i|}{|A_i| + |F_i|}
    \end{equation} \\

    Root Mean Squared Error (RMSE) & 
    \begin{equation}
        \sqrt{\frac{1}{n} \sum_{i=1}^{n} (F_i - A_i)^2}
    \end{equation} \\

    Fourth Root Mean Fourth Power Difference (FRMFPD) &
    \begin{equation}
        \left(\frac{1}{n} \sum_{i=1}^{n} (F_i - A_i)^4\right)^{\frac{1}{4}}
    \end{equation} \\

    Quantile of Absolute Percentage Error ($Q_{\alpha}(\text{APE})$) &
    \begin{equation}
        \inf \left\{ x \in \mathbb{R} : P(|\text{APE}| \leq x) \geq \alpha \right\}
    \end{equation} \\

    Wasserstein Distance between APE and Uniform dist. (WD) &
    \begin{equation}
        W = \int_{0}^{1} \left| F^{-1}_{\text{APE}}(p) - F^{-1}_{U}(p) \right| \, dp
    \end{equation} \\

    \bottomrule
    \end{tabularx}
    \vspace{0.5em}
    \emph{Note: A – actuals, F – forecasts, n – number of observations, APE – Absolute Percentage Error $= |F_i - A_i| / A_i$.}
\end{table}
\noindent The metrics in Table~\ref{tab:eval_metrics} are selected as follows. \textbf{MAPE} is the primary model selection criterion: it expresses error as a percentage of the actual price, is easily communicated to business stakeholders, and penalises under-forecasting more than over-forecasting (asymmetry in the denominator). Its known limitation is instability near zero; however, the 300~PLN price floor imposed at preprocessing makes near-zero values impossible in this dataset, so this issue does not apply here. \textbf{W-MAPE} applies 10\% symmetric Winsorisation to the distribution of absolute percentage errors before averaging, reducing the influence of extreme-error items without excluding them entirely. \textbf{SMAPE} provides a symmetric alternative: it penalises over- and under-forecasting equally by normalising by the average of actual and forecast values, making it useful for detecting directional bias. \textbf{RMSE} squares errors before averaging, giving disproportionate weight to large deviations - important for high-value items where a large absolute error has direct financial consequences. \textbf{FRMFPD} (Fourth Root Mean Fourth Power Difference) applies an even stronger penalty to extreme errors by raising residuals to the fourth power; it is used when outlier resilience of the business process is low. \textbf{$Q_{90}(\text{APE})$} reports the 90th percentile of the absolute percentage error distribution, characterising worst-case performance across items rather than average performance. Finally, the \textbf{Wasserstein Distance} between the empirical APE distribution and $\mathcal{U}[0, 0.05]$ quantifies how far the error distribution is from the business target: subscription pricing is considered operationally safe when errors are uniformly spread below 5\% - a threshold established by the company's risk management framework. MAPE is used as the primary metric throughout; the remaining metrics are reported for completeness and cross-validation.

\subsection{Experiment setting}

We evaluate forecasting performance on an aforementioned dataset containing daily prices of used electronic items scraped from major Polish marketplaces. 
Models are trained using rolling-window cross-validation described in section~\ref{sec:cross-valid}. For each forecasting horizon $h \in \{1, 7, 30, 90, 180, 365\}$ days, we generate multiple training-test splits. Each split includes a fixed-length training window followed by an out-of-sample (OOS) test segment of length $h$. The number of cross-validation folds and gap settings are adjusted per horizon (e.g., shorter training windows and more folds for shorter $h$).

Model training involves two stages. First, hyperparameter tuning is conducted using the Ray Tune framework \parencite{liaw2018tune} with a random search strategy. For each model and each OOS split, 75 candidate configurations are sampled from a predefined search space and evaluated on the inner validation windows. The searched parameters include: learning rate, batch size, maximum training steps, and input window size; architecture-specific parameters (e.g., number of layers, hidden units, and attention heads for transformer models) are also tuned (see Appendix~\ref{app:hpsearch} for the full search ranges). Second, once optimal configurations are identified per OOS split, they are aggregated (mean for continuous parameters, mode for categorical ones) and used to train a final model on the full training period (up to 30 June 2024), with 1 July 2024 marking the start of the held-out test sample.

In each OOS split, models are trained once over the training period and evaluated over the inner validation windows (see Table~\ref{tab:windows}). The best hyperparameters from each OOS split are stored and aggregated to produce the final configuration, which is then used to train the production model on the entire training history.

\begin{table}[H]
\caption{Number of windows in OOS and number of OOS in the experiment}\label{tab:windows}
\centering
\begin{tabularx}{0.5\textwidth}{r|cc}
\toprule
Horizon & Windows in OOS & No OOS \\ \midrule
1       & 30             & 30     \\
7       & 14             & 9      \\
30      & 12             & 9      \\
90      & 8              & 3      \\
180     & 8              & 3      \\
365     & 4              & 3      \\ \bottomrule
\end{tabularx}
\end{table}

All models are implemented using the \texttt{NeuralForecast} library and trained on Graphics Processing Unit (GPU) using PyTorch Lightning. The experiment supports both global (multi-series) and local (per-series) forecasting but here focuses on global forecasting using all available time series jointly. All experiments were run on a single NVIDIA GPU server. The random seed is fixed at 2023 for all trials. The full hyperparameter search ranges are listed in Appendix~\ref{app:hpsearch}.

Table~\ref{tab:training_time} summarises the per-trial training time observed during the random search, based on recorded wall-clock times from the Ray Tune logs.

\begin{table}[H]
\centering
\footnotesize
\caption{Per-trial training time during hyperparameter search (wall-clock seconds on a single GPU)}
\label{tab:training_time}
\begin{tabular}{lrrrr}
\toprule
\textbf{Model} & \textbf{Trials logged} & \textbf{Median (s)} & \textbf{Max (s)} \\
\midrule
N-BEATS   & 7{,}137  & 14  & 183  \\
N-HiTS    & 6{,}900  & 15  & 39   \\
TFT       & 29{,}567 & 9   & 333  \\
PatchTST  & 6{,}902  & 18  & 341  \\
Informer  & 13{,}350 & 86  & 2{,}447 \\
\bottomrule
\end{tabular}
\end{table}

\paragraph{Forecasting strategy.} Statistical baselines (ARIMA, ETS, Theta, Na\"{i}ve) generate multi-step forecasts recursively - producing one step at a time and feeding each prediction back as input for the next step. All deep learning models (LSTM, TCN, N-BEATS, N-HiTS, TFT, PatchTST, Informer) use a direct multi-output strategy: they predict all $h$ future steps simultaneously in a single forward pass, avoiding the compounding of recursive errors at the cost of requiring horizon-specific training.

\paragraph{Data leakage prevention.} All hyperparameter tuning is performed exclusively on out-of-sample windows that precede the held-out test period. The final test sample begins on 1 July 2024 and was never used during model selection or hyperparameter search. The training period ends on 30 June 2024 for all models and all horizons. This strict temporal separation ensures that no information from the test period influences any model configuration decision.

\subsubsection{Evaluation approach}
\label{subsec:evaluation-protocols}

\paragraph{Common setup (applies to all protocols)}
\begin{enumerate}
  \item For each series, construct rolling windows consistent with the paper’s data section (train window $\rightarrow$ forecast origin $\rightarrow$ evaluation window of length $h\in H$).
  \item Use the model- and horizon-specific best configuration selected per horizon (as defined in the \ref{subsec:evaluation-protocols} section) unless stated otherwise in Protocol 3.
  \item For each model and horizon, generate forecasts for all series and all rolling origins.
  \item Evaluate with the paper’s metrics set.
  \item Aggregate per horizon across series and origins using the paper’s aggregation rules (means for central metrics; reported quantiles for tail metrics where applicable).
\end{enumerate}

\paragraph{Protocol 1: multi-horizon (average-over-trajectory) evaluation}
\begin{enumerate}
  \item For each $h\in H$ and each model, train using the horizon-specific configuration.
  \item From each forecast origin, produce the full $h$-step trajectory.
  \item Score \emph{all} steps $t+1,\dots,t+h$ and average the stepwise errors within each trajectory.
  \item Aggregate across origins and series to obtain model-level metrics per $h$.
\end{enumerate}

\paragraph{Protocol 2: One-shot endpoint evaluation}
\begin{enumerate}
  \item For each $h\in H$ and each model, train using the horizon-specific configuration.
  \item From each forecast origin, extract only the terminal prediction at step $t+h$.
  \item Compute metrics using the endpoint errors only.
  \item Aggregate across origins and series to obtain model-level metrics per $h$.
\end{enumerate}

\paragraph{Protocol 3: Fixed far-horizon transfer (train-on-365, test-on-all)}
\begin{enumerate}
  \item For each model, \emph{train once} on the $h{=}365$ setup using its 365-day configuration.
  \item From each forecast origin, output the full 365-step forecast trajectory.
  \item For each evaluation horizon $h\in H$, slice the first $h$ steps of that single trajectory and score only step $t{+}h$.
  \item Aggregate across origins and series to obtain metrics per $(\text{model},h)$ and compare to Protocols~1--2 to assess whether a single far-horizon model can substitute for horizon-specific models.
\end{enumerate}
\paragraph{Variance analysis across top configurations}
\begin{enumerate}
  \item For each $(\text{model},h)$, collect the $K$ best hyperparameter configurations from the tuning phase (as described in Section~\ref{sec:cross-valid}), where $K\geq 5$.
  \item Compute the standard deviation across configurations for: learning rate, batch size, max training steps, input window size.
  \item Report these SDs per $(\text{model},h)$ in a summary table (as in Table~\ref{tab:hyperparam_std}); highlight horizons with notably high/low variance.
\end{enumerate}
\paragraph{Best versus average performance stability}
\begin{enumerate}
  \item For each $(\text{model},h)$, record:
        \begin{enumerate}
          \item \textbf{Best MAPE} (from the selected final configuration used in the main results).
          \item \textbf{Mean MAPE} across the $K$ tuning configurations (out-of-sample validation windows).
        \end{enumerate}
  \item Perform a paired two-sample $t$-test and a Wilcoxon signed-rank test comparing the per-window APE sequences (or MAPE across windows) of the best configuration vs.\ the $K$-config average.
  \item Report $p$-values alongside best/mean MAPE per $(\text{model},h)$ (as in Table~\ref{tab:tab5}); mark statistically significant gaps.
\end{enumerate}

\noindent The experimental design employs three complementary evaluation protocols that distinguish between trajectory fitness (Protocol 1), terminal accuracy (Protocol 2), and model reuse across horizons (Protocol 3). This separation allows us to address both scientific and operational questions, such as the suitability of models for rolling pricing versus long-term planning, and whether a single far-horizon model can replace a family of horizon-specific models. 
Horizon stratification from 1 to 365 days reflects the horizon-dependent signal-to-noise structure of the market: short horizons assess the ability to follow local trends, while long horizons test the capacity to extrapolate structural changes and regime shifts. Robustness analysis ensures that the conclusions are not artifacts of hyperparameter tuning, with variance measures and best-versus-typical performance gaps measuring both tunability risk and deployment stability.

\section{Results}

We first present the model metrics calculated over the entire forecasting horizon, with errors evaluated for each timestep in the prediction period (table \ref{tab:tab1}) and then targetted to predict exactly the last observation from the horizon (one-shot prediction, table \ref{tab:tab2}).

Table \ref{tab:tab1} presents the comparative performance of the tested forecasting models evaluated across all prediction horizons. At the shortest horizon (1 day), nearly all models (except Informer) exhibit near-identical error rates (MAPE $\approx$ 0.0072), with NHITS, PatchTST, and TCN marginally outperforming the others across multiple metrics. Informer lags behind with a noticeably higher error (MAPE $\approx$ 0.0102).

As the forecast horizon increases to 7 and 30 days, model differentiation becomes more pronounced. NHITS and NBEATS maintain competitive performance, with PatchTST and TFT also showing strong results. Informer and Theta gradually underperform, especially beyond 30 days.

At horizons of 90 and 180 days, performance degradation is substantial across all models. While NHITS and NBEATS still rank among the top performers (MAPE $\approx$ 0.049 at 90 days, $\approx$ 0.067 at 180 days), traditional models such as ETS and Theta show higher MAPE and RMSE values. Informer and LSTM display particularly poor generalization, with LSTM reaching MAPE $\approx$ 0.162 at 180 days and $\approx$ 0.372 at 365 days.

At the 365-day horizon, deep learning models remain competitive only in the example of NBEATS, all other models exhibit significant error growth. The naive baseline performs comparably to most machine learning models, highlighting the inherent difficulty of long-range forecasting in this domain.

In terms of FRMFPD and Wasserstein distance, performance trends mirror those of the traditional error metrics, reinforcing the conclusion that NHITS, PatchTST, and NBEATS consistently offer the most robust results across forecast horizons.

The comparative results across all forecast horizons reveal consistent differences between traditional statistical models (ARIMA, ETS, Theta) and modern deep learning architectures (NHITS, PatchTST, NBEATS, TFT). At short horizons (1–7 days), performance differences are marginal; all models -- including traditional approaches -- achieve near-identical metrics across MAPE, RMSE, and distributional scores such as Wasserstein distance. Interestingly, the naive model performs competitively at short horizons, indicating that the liquidity of this market is rather small, because the temporal changes are rather small over short time frames. However, as the forecast horizon extends (30–365 days), deep learning models begin to show a distinct advantage. For instance, at 180 days, NHITS and NBEATS maintain relatively low MAPE values ($\approx$0.0675–0.0678), while traditional methods such as Theta and ETS reach significantly higher errors ($\approx$0.0939 and $\approx$0.0919, respectively). The differences between models are small, first because most of them produce aggregated results that are averaged, and second because day-to-day price fluctuations are relatively minor, making even a naive forecast reasonably effective. At the 365-day horizon, this gap widens further: deep models like NBEATS sustain competitive accuracy (MAPE $\approx$ 0.0851), whereas Theta and ETS degrade to MAPE values above 0.14. Notably, the naive model also performs comparably to many statistical and deep learning models at this long horizon (MAPE = 0.1751), underscoring the challenge of long-range forecasting and the limited incremental value of complex architectures in highly uncertain regimes. These findings indicate that while statistical models can approximate short-term dynamics well, they struggle to capture long-term nonlinearities and structural shifts present in the used electronics market.

\newgeometry{left=0.1cm,bottom=0.1cm, right=0.1cm,top=0.1cm}

\begin{table}[p]
\footnotesize
\caption{Metric results over whole horizon for the tested models with all tested forecast horizons}\label{tab:tab1}
\centering
\begin{tabularx}{0.8\linewidth}{cr|ccccccc}
\toprule
Horizon               & Model     & MAPE   & W-MAPE & SMAPE  & RMSE      & FRMFPD    & Q90(APE) & Wasserstein (APE, U{[}0,5\%{]}) \\ \midrule
\multirow{11}{*}{1}   & NHITS     & \textbf{0.0072} & \textbf{0.0055} & \textbf{0.0072} & 45.7661   & 102.7146  & 0.0191   & 0.0183                         \\
                      & PatchTST  & 0.0072 & 0.0055 & 0.0072 & 45.7619   & 102.7242  & 0.0191   & 0.0185                         \\
                      & Naive     & 0.0072 & 0.0055 & 0.0072 & 45.7619   & 102.7242  & 0.0191   & 0.0194                         \\
                      & TCN       & 0.0072 & 0.0055 & 0.0072 & 45.7617   & 102.7242  & 0.0191   & 0.0188                         \\
                      & LSTM      & 0.0072 & 0.0055 & 0.0072 & 45.7615   & 102.7241  & 0.0192   & 0.0182                         \\
                      & NBEATS    & 0.0072 & 0.0055 & 0.0072 & 45.9056   & 102.697   & \textbf{0.0186}   & 0.0193                         \\
                      & TFT       & 0.0072 & 0.0055 & 0.0072 & \textbf{45.7566}   & 102.7235  & 0.0192   & 0.0189                         \\
                      & ETS   & 0.0073 & 0.0056 & 0.0073 & 45.6058   & 101.7074  & 0.0193   & 0.018                          \\
                      & Theta & 0.0073 & 0.0056 & 0.0073 & 46.1668   & 102.379   & 0.019    & 0.0176                         \\
                      & ARIMA & 0.0074 & 0.0057 & 0.0074 & 45.7468   & \textbf{97.2374}   & 0.0192   & 0.0173                         \\
                      & Informer  & 0.0102 & 0.008  & 0.0102 & 62.4289   & 135.4354  & 0.0223   & \textbf{0.015}                          \\ \midrule
\multirow{11}{*}{7}   & NHITS     & \textbf{0.0155} & \textbf{0.0117} & \textbf{0.0155} & 100.6955  & 218.1019  & 0.04     & 0.0155                         \\
                      & TFT       & 0.0155 & 0.0117 & 0.0155 & 100.696   & 218.1034  & 0.04     & 0.0155                         \\
                      & Naive     & 0.0155 & 0.0117 & 0.0155 & \textbf{100.6952}  & 218.1003  & 0.04     & 0.0159                         \\
                      & PatchTST  & 0.0155 & 0.0118 & 0.0155 & 100.8674  & 218.4717  & 0.04     & 0.0158                         \\
                      & NBEATS    & 0.0157 & 0.0119 & 0.0157 & 103.5587  & 224.9511  & 0.041    & 0.0157                         \\
                      & ARIMA & 0.0159 & 0.0122 & 0.0159 & 101.3245  & \textbf{217.877}  & 0.0409   & 0.015                          \\
                      & ETS   & 0.016  & 0.0123 & 0.016  & 100.8568  & 217.9035  & 0.0411   & 0.0152                         \\
                      & Theta & 0.0162 & 0.0125 & 0.0162 & 102.9733  & 220.8115  & 0.0412   & 0.0151                         \\
                      & Informer  & 0.0166 & 0.0129 & 0.0166 & 109.0882  & 231.9855  & \textbf{0.0391}   & 0.0144                         \\
                      & TCN       & 0.0177 & 0.0137 & 0.0178 & 115.4247  & 246.5581  & 0.0406   & \textbf{0.0138}                         \\
                      & LSTM      & 0.0177 & 0.0137 & 0.0178 & 115.483   & 246.7519  & 0.0406   & 0.0143                         \\ \midrule
\multirow{11}{*}{30}  & NBEATS    & \textbf{0.0333} & 0.0273 & \textbf{0.0337} & \textbf{220.1533}  & 457.1854  & \textbf{0.0854}   & 0.017                          \\
                      & NHITS     & 0.0335 & \textbf{0.0269} & 0.0338 & 225.6938  & 467.3998  & 0.0856   & 0.0184                         \\
                      & PatchTST  & 0.0344 & 0.0283 & 0.0347 & 220.8047  & 458.4318  & 0.0846   & \textbf{0.0165}                         \\
                      & ARIMA & 0.0353 & 0.0286 & 0.0352 & 222.5629  & 456.0626  & 0.0877   & 0.0175                         \\
                      & TFT       & 0.0353 & 0.0286 & 0.0352 & 222.458   & 455.7388  & 0.0865   & 0.0172                         \\
                      & Naive     & 0.0353 & 0.0286 & 0.0352 & 222.4574  & \textbf{455.7312}  & 0.0865   & 0.0169                         \\
                      & ETS   & 0.0356 & 0.029  & 0.0356 & 229.6427  & 470.2119  & 0.0901   & 0.0181                         \\
                      & Informer  & 0.0368 & 0.0299 & 0.0368 & 231.0165  & 484.8276  & 0.0873   & 0.0172                         \\
                      & Theta & 0.0379 & 0.0311 & 0.0381 & 233.7201  & 470.6374  & 0.0896   & 0.0168                         \\
                      & LSTM      & 0.0469 & 0.039  & 0.0469 & 294.1656  & 628.5077  & 0.1058   & 0.022                          \\
                      & TCN       & 0.047  & 0.0391 & 0.0469 & 294.1793  & 628.2299  & 0.1059   & 0.0218                         \\ \midrule
\multirow{11}{*}{90}  & NHITS     & \textbf{0.0489} & \textbf{0.0403} & \textbf{0.0495} & 334.7708  & 767.6299  & 0.1199   & \textbf{0.0252}                         \\
                      & NBEATS    & 0.049  & 0.0404 & 0.0495 & \textbf{334.3013}  & 767.3287  & 0.1197   & 0.0252                         \\
                      & PatchTST  & 0.0559 & 0.0481 & 0.056  & 334.6301  & \textbf{750.2452 } & \textbf{0.1185}   & 0.0311                         \\
                      & ARIMA & 0.0575 & 0.048  & 0.0566 & 343.5917  & 759.6736  & 0.1259   & 0.0327                         \\
                      & ETS   & 0.059  & 0.0498 & 0.0584 & 359.9688  & 774.9137  & 0.1317   & 0.0342                         \\
                      & TFT       & 0.0592 & 0.0498 & 0.058  & 343.2953  & 759.3064  & 0.1204   & 0.0343                         \\
                      & Naive     & 0.0592 & 0.0498 & 0.058  & 343.2958  & 759.3065  & 0.1204   & 0.0342                         \\
                      & Informer  & 0.0614 & 0.0511 & 0.0633 & 370.9433  & 778.5492  & 0.1344   & 0.0364                         \\
                      & Theta & 0.0621 & 0.053  & 0.062  & 362.4439  & 764.9393  & 0.1367   & 0.0371                         \\
                      & LSTM      & 0.0903 & 0.0813 & 0.0886 & 422.7921  & 867.1583  & 0.1547   & 0.0655                         \\
                      & TCN       & 0.0906 & 0.0815 & 0.0889 & 423.4253  & 868.0785  & 0.1555   & 0.0656                         \\ \midrule
\multirow{11}{*}{180} & NHITS     & \textbf{0.0675} & \textbf{0.0554} & \textbf{0.0683} & \textbf{381.5329}  & 855.7468  & \textbf{0.16}     & \textbf{0.0426}                         \\
                      & NBEATS    & 0.0678 & 0.0558 & 0.0692 & 389.308   & 869.4171  & 0.1632   & 0.0429                         \\
                      & ARIMA & 0.0899 & 0.0786 & 0.0855 & 397.4302  & \textbf{820.2355}  & 0.1968   & 0.0651                         \\
                      & ETS   & 0.0919 & 0.0806 & 0.0881 & 423.8902  & 843.8461  & 0.2001   & 0.0669                         \\
                      & Theta & 0.0939 & 0.0817 & 0.0919 & 445.4832  & 868.5466  & 0.2069   & 0.069                          \\
                      & PatchTST  & 0.0941 & 0.0832 & 0.0886 & 400.3098  & 820.4337  & 0.1966   & 0.0692                         \\
                      & TFT       & 0.0942 & 0.0833 & 0.0887 & 400.3778  & 820.4506  & 0.1968   & 0.0692                         \\
                      & Naive     & 0.0942 & 0.0833 & 0.0887 & 400.378   & 820.4506  & 0.1968   & 0.0691                         \\
                      & Informer  & 0.0953 & 0.0845 & 0.0896 & 406.9473  & 830.3144  & 0.1972   & 0.0703                         \\
                      & TCN       & 0.1601 & 0.1539 & 0.1471 & 561.6889  & 982.2655  & 0.288    & 0.1351                         \\
                      & LSTM      & 0.1624 & 0.1562 & 0.1489 & 568.3     & 990.8905  & 0.2913   & 0.1375                         \\ \midrule
\multirow{11}{*}{365} & NBEATS    & \textbf{0.0851} & \textbf{0.0739} & \textbf{0.0876} & \textbf{421.3005}  & \textbf{1040.8495} & \textbf{0.1822}   & \textbf{0.06}                           \\
                      & Theta & 0.1494 & 0.1312 & 0.1424 & 754.371   & 1692.3416 & 0.3614   & 0.1244                         \\
                      & ETS   & 0.1621 & 0.1525 & 0.1558 & 696.3389  & 1597.4075 & 0.3517   & 0.1371                         \\
                      & TFT       & 0.1751 & 0.168  & 0.1578 & 721.6348  & 1635.9834 & 0.3693   & 0.1501                         \\
                      & Naive     & 0.1751 & 0.168  & 0.1578 & 721.6388  & 1635.9858 & 0.3694   & 0.1501                         \\
                      & NHITS     & 0.1758 & 0.1691 & 0.1587 & 715.6719  & 1617.5776 & 0.3683   & 0.1508                         \\
                      & ARIMA & 0.1771 & 0.1649 & 0.1699 & 740.3997  & 1649.4053 & 0.3774   & 0.1521                         \\
                      & PatchTST  & 0.1789 & 0.1721 & 0.161  & 723.0339  & 1622.0771 & 0.3733   & 0.1539                         \\
                      & Informer  & 0.1809 & 0.1741 & 0.1625 & 740.1586  & 1682.16   & 0.3752   & 0.156                          \\
                      & LSTM      & 0.372  & 0.351  & 0.2999 & 1337.5816 & 2554.7848 & 0.7093   & 0.3469                         \\
                      & TCN       & 0.3729 & 0.3519 & 0.3004 & 1339.8503 & 2554.3506 & 0.7107   & 0.3479                         \\ \bottomrule
\end{tabularx}
\end{table}

\restoregeometry

\begin{figure}[H]
    \centering
    \includegraphics[width=\linewidth]{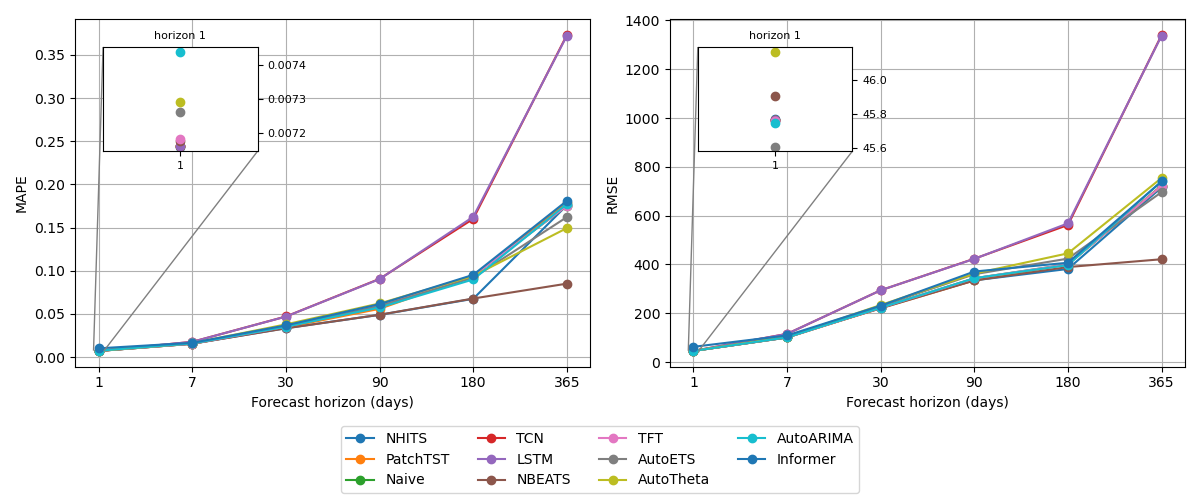}
    \caption{MAPE and RMSE results for all tested models with all forecast horizons}
    \label{fig:metrics-all}
\end{figure}

\begin{figure}[H]
    \centering
        \caption{Forecasts for exemplary items for forecast horizon = 1}

    \includegraphics[width=1\linewidth]{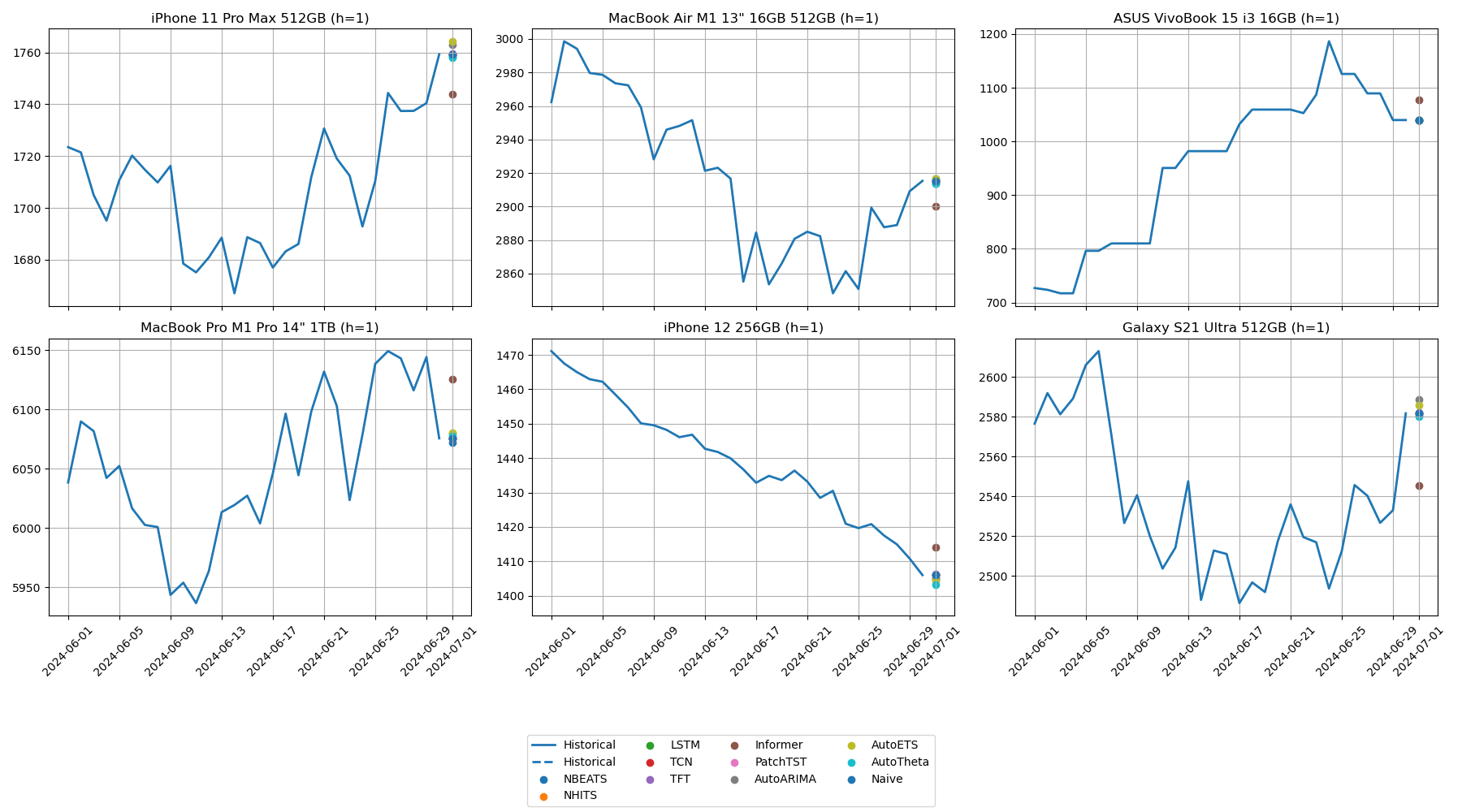}
    \label{fig:fig2}
\end{figure}

\begin{figure}[H]
    \centering
        \caption{Forecasts for exemplary items for forecast horizon = 365}

    \includegraphics[width=1\linewidth]{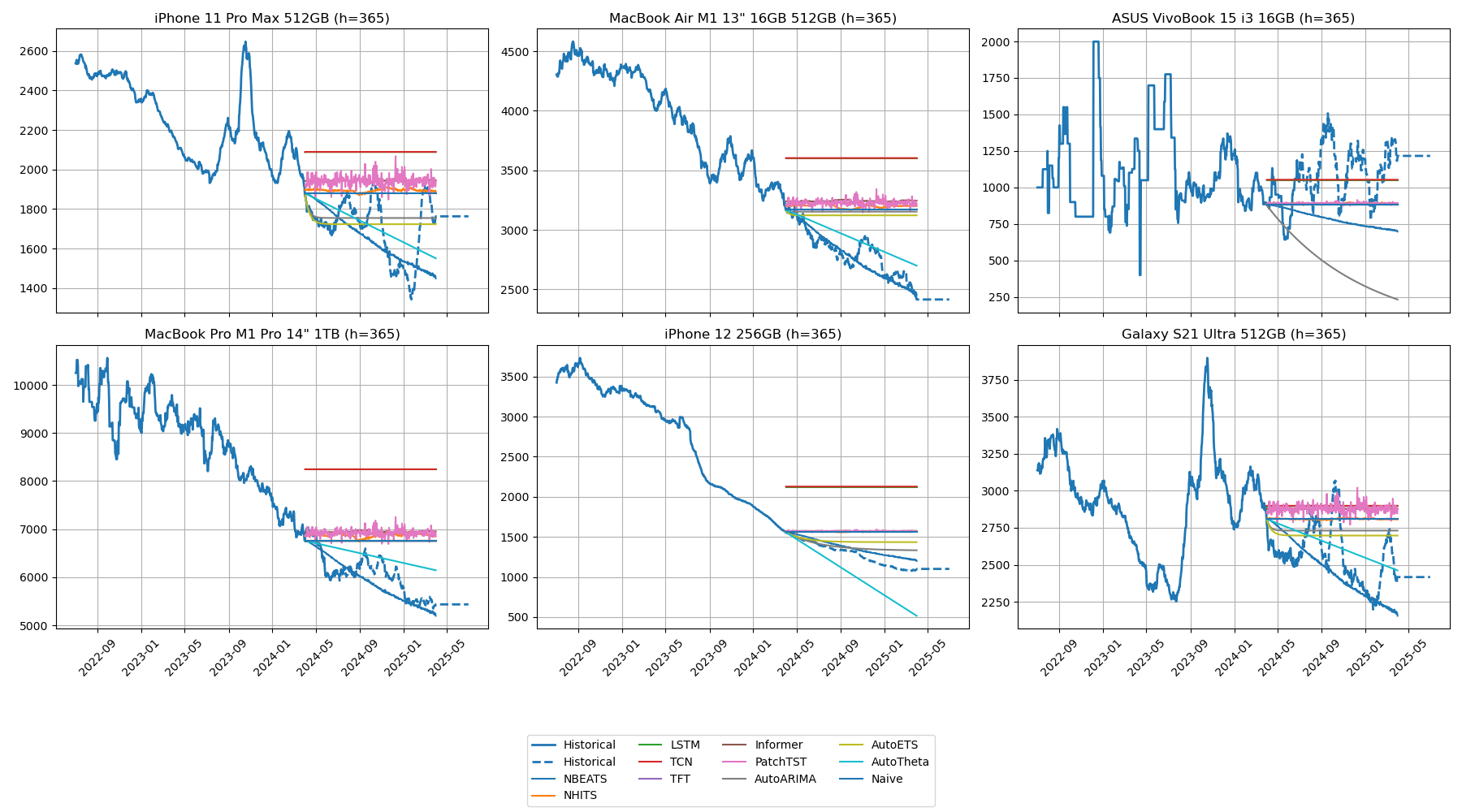}
    \label{fig:fig3}
\end{figure}

Figure~\ref{fig:metrics-all} visualises the aggregated error metrics - MAPE and RMSE - across forecast horizons (1 to 365 days) for all evaluated models. The plots confirm the earlier findings: errors remain low and tightly clustered at short horizons, then diverge significantly over time. N-HiTS, PatchTST, TFT, and N-BEATS demonstrate lower error growth rates, while TCN and LSTM suffer pronounced degradation.

Figure~\ref{fig:fig2} illustrates 1-day-ahead forecasts for six selected items. At this short horizon, most models yield accurate forecasts closely aligned with recent trends. Informer and LSTM occasionally produce noticeable deviations from observed values. Figure~\ref{fig:fig3} presents the 365-day-ahead forecasts for the same items, highlighting the challenge of long-horizon prediction.

The poor performance of LSTM and TCN at this horizon requires explicit explanation. Both models produce flat or strongly diverging trajectories that are detached from the underlying downward depreciation trend visible in the historical data. This failure has two root causes. First, both architectures overfit to short-term noise during training: the rolling loss optimisation favours patterns that explain day-to-day fluctuations, which do not generalise to year-long structural change. Second, neither model incorporates explicit structural decomposition - they have no mechanism to separately model and extrapolate the long-term trend component, so when the input window context is insufficient relative to the horizon length, they effectively revert to a flat or random-walk prediction. The visible divergence in some LSTM and TCN curves is caused by accumulated recursive prediction errors in configurations where the flat initial forecast drifts away from the actual depreciation path.

N-BEATS, by contrast, leverages its residual stacking to progressively subtract estimated trend and seasonality components from the input, producing forecasts that track the structural price trajectory even at this extended horizon. N-HiTS and PatchTST also follow extrapolated trends more closely than recurrent models, though all methods show increasing uncertainty beyond 90 days.

\newgeometry{left=0.1cm,bottom=0.1cm, right=0.1cm,top=0.1cm}
\begin{table}[p]
\footnotesize
\centering
\caption{Metric results for point forecast at horizon end for the tested models with all tested forecast horizons}\label{tab:tab2}
\begin{tabularx}{0.8\linewidth}{cr|ccccccc}

\toprule
Horizon               & Model     & MAPE   & W-MAPE & SMAPE  & RMSE      & FRMFPD    & Q90(APE) & Wasserstein(APE, U{[}0,5\%{]}) \\ \midrule
\multirow{11}{*}{1}   & NHITS     & \textbf{0.0072} & \textbf{0.0055} &\textbf{ 0.0072} & 45.7661   & 102.7146  & 0.0191   & 0.0188                         \\
                      & PatchTST  & 0.0072 & 0.0055 & 0.0072 & 45.7619   & 102.7242  & 0.0191   & \textbf{0.0173}                         \\
                      & Naive     & 0.0072 & 0.0055 & 0.0072 & 45.7619   & 102.7242  & 0.0191   & 0.0193                         \\
                      & TCN       & 0.0072 & 0.0055 & 0.0072 & 45.7617   & 102.7242  & 0.0191   & 0.0187                         \\
                      & LSTM      & 0.0072 & 0.0055 & 0.0072 & 45.7615   & 102.7241  & 0.0192   & 0.0189                         \\
                      & NBEATS    & 0.0072 & 0.0055 & 0.0072 & 45.9056   & 102.697   & \textbf{0.0186}   & 0.0176                         \\
                      & TFT       & 0.0072 & 0.0055 & 0.0072 & 45.7566   & 102.7235  & 0.0192   & 0.0179                         \\
                      & ETS   & 0.0073 & 0.0056 & 0.0073 & \textbf{45.6058}   & 101.7074  & 0.0193   & 0.0183                         \\
                      & Theta & 0.0073 & 0.0056 & 0.0073 & 46.1668   & 102.379   & 0.019    & 0.018                          \\
                      & ARIMA & 0.0074 & 0.0057 & 0.0074 & 45.7468   & \textbf{97.2374}   & 0.0192   & 0.0182                         \\
                      & Informer  & 0.0102 & 0.008  & 0.0102 & 62.4289   & 135.4354  & 0.0223   & 0.0163                         \\ \midrule
\multirow{11}{*}{7}   & Informer  & \textbf{0.0232} & \textbf{0.0186} & \textbf{0.0233} & \textbf{146.3065}  & 285.6409  & \textbf{0.0602}   & 0.015                          \\
                      & TFT       & 0.0237 & 0.0191 & 0.0238 & 146.6123  & 271.7396  & 0.063    & 0.0153                         \\
                      & NHITS     & 0.0237 & 0.0191 & 0.0238 & 146.6115  & 271.7376  & 0.063    & 0.0151                         \\
                      & Naive     & 0.0237 & 0.0191 & 0.0238 & 146.6108  & 271.7355  & 0.063    & 0.0154                         \\
                      & PatchTST  & 0.0238 & 0.0191 & 0.0238 & 147.1546  & 272.7552  & 0.0633   & 0.0153                         \\
                      & ARIMA & 0.0239 & 0.0194 & 0.024  & 147.5068  & \textbf{271.271}   & 0.0648   & 0.0153                         \\
                      & NBEATS    & 0.0241 & 0.0197 & 0.0242 & 153.5821  & 284.341   & 0.0677   & 0.0147                         \\
                      & TCN       & 0.0244 & 0.0196 & 0.0246 & 151.2813  & 298.5265  & 0.0609   & 0.0144                         \\
                      & LSTM      & 0.0244 & 0.0196 & 0.0246 & 151.3896  & 298.7997  & 0.061    & \textbf{0.0133}                         \\
                      & ETS   & 0.0245 & 0.0205 & 0.0246 & 147.6659  & 271.9561  & 0.0691   & 0.0147                         \\
                      & Theta & 0.025  & 0.0206 & 0.0251 & 151.1309  & 277.644   & 0.0654   & 0.0137                         \\ \midrule
\multirow{11}{*}{30}  & NBEATS    & \textbf{0.0519} & 0.0439 & \textbf{0.0531} & 330.516   & 632.0378  & 0.127    & \textbf{0.0278}                         \\
                      & NHITS     & 0.0532 & \textbf{0.0437} & 0.0541 & 335.4754  & 627.706   & 0.1269   & 0.0297                         \\
                      & PatchTST  & 0.0556 & 0.0479 & 0.0564 & 329.4379  & 620.7193  & 0.1296   & 0.0304                         \\
                      & ARIMA & 0.0577 & 0.0488 & 0.0578 & 329.0414  & 609.8773  & 0.1297   & 0.0326                         \\
                      & TFT       & 0.058  & 0.0488 & 0.0579 & \textbf{329.0072}  & 609.253   & 0.1273   & 0.0325                         \\
                      & Naive     & 0.058  & 0.0488 & 0.0579 & 329.0076  & \textbf{609.2439}  & 0.1273   & 0.0324                         \\
                      & ETS   & 0.058  & 0.0482 & 0.0582 & 341.8165  & 631.4364  & 0.1282   & 0.033                          \\
                      & Informer  & 0.0586 & 0.0492 & 0.0586 & 334.7129  & 641.3583  & 0.1262   & 0.033                          \\
                      & Theta & 0.0631 & 0.0525 & 0.0636 & 348.2045  & 630.769   & \textbf{0.1248}   & 0.0381                         \\
                      & TCN       & 0.0639 & 0.0548 & 0.0641 & 375.867   & 751.8929  & 0.1259   & 0.0381                         \\
                      & LSTM      & 0.064  & 0.0549 & 0.0641 & 375.2452  & 748.229   & 0.1261   & 0.0411                         \\ \midrule
\multirow{11}{*}{90}  & NHITS     & \textbf{0.0713} & \textbf{0.062}  & \textbf{0.0729} & 419.3905  & 910.375   & 0.1692   & \textbf{0.0464}                         \\
                      & NBEATS    & 0.0713 & 0.062  & 0.0731 & 422.9807  & 917.8742  & 0.1688   & 0.0465                         \\
                      & PatchTST  & 0.0829 & 0.0748 & 0.0823 & \textbf{375.171}   & \textbf{810.6204}  & \textbf{0.1514}   & 0.0581                         \\
                      & ARIMA & 0.0876 & 0.0761 & 0.0855 & 401.7592  & 850.9095  & 0.1614   & 0.0619                         \\
                      & TFT       & 0.0894 & 0.0784 & 0.0866 & 401.309   & 850.9294  & 0.1549   & 0.0643                         \\
                      & Naive     & 0.0894 & 0.0784 & 0.0866 & 401.3098  & 850.9296  & 0.1549   & 0.0647                         \\
                      & Informer  & 0.091  & 0.0819 & 0.0942 & 438.4633  & 848.5662  & 0.2024   & 0.0664                         \\
                      & ETS   & 0.0915 & 0.0803 & 0.0902 & 435.1076  & 875.4418  & 0.1769   & 0.0673                         \\
                      & TCN       & 0.0939 & 0.0849 & 0.0945 & 393.2161  & 819.1334  & 0.1543   & 0.0694                         \\
                      & Theta & 0.0959 & 0.0855 & 0.0962 & 445.2963  & 868.3187  & 0.1891   & 0.0709                         \\
                      & LSTM      & 0.1074 & 0.0986 & 0.1036 & 437.5474  & 865.9463  & 0.1683   & 0.082                          \\ \midrule
\multirow{11}{*}{180} & NBEATS    & \textbf{0.0847} & \textbf{0.0707} & 0.0899 & 398.1981  & 936.4066  & 0.1907   & \textbf{0.0594}                         \\
                      & NHITS     & 0.0855 & 0.0718 & \textbf{0.0891} & \textbf{386.2918}  & \textbf{914.6865}  & \textbf{0.1863}   & 0.0614                         \\
                      & Theta & 0.1474 & 0.1347 & 0.1446 & 609.4986  & 1110.9264 & 0.3028   & 0.1233                         \\
                      & ARIMA & 0.1537 & 0.1419 & 0.1446 & 550.4262  & 1026.6386 & 0.2492   & 0.1295                         \\
                      & ETS   & 0.1559 & 0.1451 & 0.1483 & 572.6912  & 1038.7284 & 0.2549   & 0.1295                         \\
                      & PatchTST  & 0.1638 & 0.1534 & 0.1516 & 558.6546  & 1027.0845 & 0.2471   & 0.1398                         \\
                      & TFT       & 0.164  & 0.1535 & 0.1517 & 558.7997  & 1027.1511 & 0.2472   & 0.1387                         \\
                      & Naive     & 0.164  & 0.1535 & 0.1517 & 558.8     & 1027.151  & 0.2472   & 0.1401                         \\
                      & Informer  & 0.1649 & 0.1542 & 0.1522 & 567.9022  & 1045.9629 & 0.2464   & 0.1401                         \\
                      & TCN       & 0.2429 & 0.2409 & 0.2153 & 784.0297  & 1297.4626 & 0.369    & 0.2181                         \\
                      & LSTM      & 0.2453 & 0.2434 & 0.2171 & 791.9544  & 1310.5741 & 0.374    & 0.2207                         \\ \midrule
\multirow{11}{*}{365} & NBEATS    & \textbf{0.1273} & \textbf{0.1148} & \textbf{0.1374} & \textbf{489.496}   & \textbf{1166.3903} & \textbf{0.2652}   & \textbf{0.1024}                         \\
                      & Theta & 0.2817 & 0.25   & 0.2668 & 1185.6031 & 2332.1161 & 0.5587   & 0.2573                         \\
                      & NHITS     & 0.2965 & 0.2932 & 0.2557 & 1082.1386 & 2184.3453 & 0.4807   & 0.2723                         \\
                      & TFT       & 0.2992 & 0.2958 & 0.2581 & 1100.5356 & 2216.6987 & 0.4777   & 0.2742                         \\
                      & Naive     & 0.2992 & 0.2958 & 0.2581 & 1100.5402 & 2216.7017 & 0.4777   & 0.2731                         \\
                      & PatchTST  & 0.3023 & 0.2993 & 0.2599 & 1098.5644 & 2184.1276 & 0.49     & 0.2775                         \\
                      & ETS   & 0.303  & 0.293  & 0.299  & 1084.7867 & 2171.9888 & 0.5174   & 0.2775                         \\
                      & Informer  & 0.3051 & 0.3017 & 0.2614 & 1127.1119 & 2275.0076 & 0.4953   & 0.2801                         \\
                      & ARIMA & 0.3167 & 0.2999 & 0.305  & 1155.6358 & 2245.3209 & 0.5075   & 0.2905                         \\
                      & LSTM      & 0.5038 & 0.5024 & 0.391  & 1703.8227 & 3016.9537 & 0.895    & 0.479                          \\
                      & TCN       & 0.5048 & 0.5034 & 0.3915 & 1706.6342 & 3019.8102 & 0.8964   & 0.4806                         \\ \bottomrule
\end{tabularx}
\end{table}

\restoregeometry

The one-shot forecasting results presented in Table \ref{tab:tab2} deviate notably from the trends observed in the whole horizon evaluation. While deep learning models such as NHITS, PatchTST, and NBEATS previously dominated across average horizons, their relative advantage diminishes in the one-shot setting - particularly at longer horizons.

At short horizons (e.g., h = 1 and h = 7), all models again exhibit near-identical performance, with MAPE values converging around 0.0072–0.0074 at h = 1. The naive model remains surprisingly competitive, closely matching the performance of both deep learning and traditional statistical models across all metrics. At intermediate horizons (e.g., h = 30 and h = 90), the ranking begins to shift. While NBEATS and NHITS still offer strong performance (MAPE $\approx$ 0.053), the uplift over traditional statistical baselines - such as ARIMA, ETS, and Theta - narrows. Additionally, models like TFT and the naive approach begin to perform comparably, particularly in RMSE and distributional metrics.

\begin{figure}[H]
    \centering
        \caption{MAPE and RMSE results for all tested models with all forecast horizons for one-shot comparisons}

    \includegraphics[width=\linewidth]{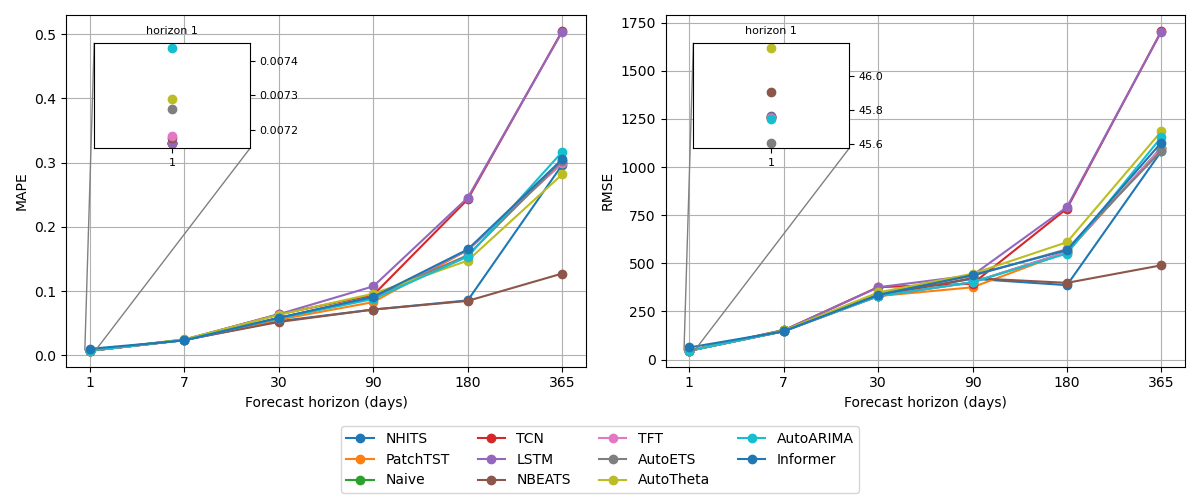}
    \label{fig:fig4}
\end{figure}

The most significant divergence occurs at the longest horizons (h = 180 and h = 365). Previously top-performing models such as NHITS and PatchTST experience pronounced performance degradation in the one-shot setting: NHITS reaches MAPE = 0.2965 at h = 365, substantially worse than its MAPE $\approx$ 0.1758 in the average-horizon evaluation. In contrast, NBEATS retains its relative strength (MAPE = 0.1273 vs MAPE = 0.0851 over whole horizon), outperforming all other models, including both deep and statistical baselines. Traditional statistical models - particularly Theta and ETS - exhibit moderate performance degradation, but remain competitive compared to most deep models, with MAPE in the 0.28–0.30 range. Notably, the naive model performs on par with TFT, NHITS, and ETS at this horizon (MAPE $\approx$ 0.2992). These results are confirmed in Figure \ref{fig:fig4}.

\newgeometry{left=0.1cm,bottom=0.1cm, right=0.1cm,top=0.1cm}

\begin{table}[p]
\footnotesize
\centering
\caption{Metric results for the tested models trained to predict 365-day horizon tested at all horizon lengths horizon end}\label{tab:tab3}
\begin{tabularx}{0.8\linewidth}{cr|ccccccc}
\toprule
Horizon               & Model     & MAPE   & W-MAPE & SMAPE  & RMSE      & FRMFPD    & Q90(APE) & Wasserstein(APE, U{[}0,5\%{]}) \\ \midrule
\multirow{11}{*}{1}   & ETS   & \textbf{0.0066} & \textbf{0.0052} & \textbf{0.0067} & \textbf{40.9375}   & 103.7432  & 0.0164   & 0.0181                         \\
                      & ARIMA & 0.0068 & 0.0053 & 0.0068 & 41.4634   & 110.2254  & \textbf{0.0157}   & 0.0175                         \\
                      & Theta & 0.0071 & 0.0057 & 0.0071 & 41.5503   & 103.5205  & 0.0168   & 0.0174                         \\
                      & TFT       & 0.0072 & 0.0058 & 0.0072 & 41.5557   & 102.996   & 0.017    & 0.0173                         \\
                      & Naive     & 0.0072 & 0.0058 & 0.0072 & 41.5559   & \textbf{102.9942}  & 0.017    & 0.018                          \\
                      & NBEATS    & 0.0072 & 0.0059 & 0.0073 & 40.2122   & 105.1397  & 0.0173   & 0.0189                         \\
                      & NHITS     & 0.0136 & 0.0114 & 0.0137 & 99.0186   & 220.4944  & 0.0316   & 0.0141                         \\
                      & PatchTST  & 0.0155 & 0.0135 & 0.0155 & 91.9994   & 203.7296  & 0.0342   & 0.0128                         \\
                      & Informer  & 0.0203 & 0.0174 & 0.0202 & 115.2189  & 243.4973  & 0.0429   & \textbf{0.0096}                         \\
                      & LSTM      & 0.1991 & 0.1812 & 0.1768 & 919.6806  & 2371.6816 & 0.3443   & 0.175                          \\
                      & TCN       & 0.2    & 0.1819 & 0.1772 & 920.8675  & 2361.1528 & 0.3455   & 0.1761                         \\ \midrule
\multirow{11}{*}{7}   & ETS   & \textbf{0.0239} & \textbf{0.0199} & \textbf{0.0238} & \textbf{121.366}   & \textbf{235.2749}  & 0.0574   & 0.0124                         \\
                      & ARIMA & 0.0248 & 0.0206 & 0.0247 & 124.8053  & 247.1408  & \textbf{0.0565}   & \textbf{0.0105}                         \\
                      & Theta & 0.0261 & 0.0225 & 0.0259 & 129.2841  & 249.4034  & 0.0648   & 0.0106                         \\
                      & NBEATS    & 0.0262 & 0.0224 & 0.026  & 127.4528  & 257.2112  & 0.0627   & 0.0107                         \\
                      & TFT       & 0.0269 & 0.0232 & 0.0267 & 127.0824  & 247.9002  & 0.0637   & 0.0109                         \\
                      & Naive     & 0.0269 & 0.0232 & 0.0267 & 127.0838  & 247.9023  & 0.0637   & 0.0108                         \\
                      & NHITS     & 0.0299 & 0.0264 & 0.0297 & 135.8043  & 251.2136  & 0.0668   & 0.0089                         \\
                      & Informer  & 0.0356 & 0.0318 & 0.0351 & 161.893   & 306.8798  & 0.0804   & 0.0119                         \\
                      & PatchTST  & 0.0442 & 0.0384 & 0.0439 & 212.618   & 402.4007  & 0.0993   & 0.0199                         \\
                      & LSTM      & 0.2065 & 0.1886 & 0.1826 & 925.1719  & 2295.0572 & 0.3552   & 0.1829                         \\
                      & TCN       & 0.2073 & 0.1892 & 0.1831 & 926.444   & 2289.1307 & 0.3562   & 0.1816                         \\ \midrule
\multirow{11}{*}{30}  & NBEATS    & \textbf{0.063}  & 0.051  & \textbf{0.0598} & \textbf{445.5976}  & \textbf{1082.4931} & 0.1461   & \textbf{0.0375}                         \\
                      & ETS   & 0.0641 & \textbf{0.0504} & 0.0605 & 455.7874  & 1119.2011 & 0.1393   & 0.0392                         \\
                      & Theta & 0.0683 & 0.0564 & 0.0647 & 476.4174  & 1141.8851 & 0.1533   & 0.0441                         \\
                      & ARIMA & 0.0689 & 0.0557 & 0.0652 & 455.534   & 1123.4636 & 0.1393   & 0.0434                         \\
                      & TFT       & 0.0725 & 0.0591 & 0.0683 & 468.5735  & 1148.5878 & 0.1389   & 0.0482                         \\
                      & Naive     & 0.0725 & 0.0591 & 0.0683 & 468.5753  & 1148.5899 & 0.1389   & 0.0476                         \\
                      & NHITS     & 0.0729 & 0.0614 & 0.0692 & 455.6636  & 1133.5601 & 0.143    & 0.0489                         \\
                      & PatchTST  & 0.0745 & 0.0622 & 0.0703 & 453.0286  & 1133.4227 & 0.14     & 0.0491                         \\
                      & Informer  & 0.0758 & 0.0624 & 0.0711 & 463.122   & 1176.4978 & \textbf{0.1364}   & 0.0507                         \\
                      & LSTM      & 0.2487 & 0.2198 & 0.2086 & 1029.0313 & 2376.7333 & 0.3944   & 0.2245                         \\
                      & TCN       & 0.2496 & 0.2206 & 0.2092 & 1030.6316 & 2373.8381 & 0.3944   & 0.2249                         \\ \midrule
\multirow{11}{*}{90}  & NBEATS    & \textbf{0.0728} & \textbf{0.06}   & \textbf{0.0731} & \textbf{391.2679}  & \textbf{827.0932}  & \textbf{0.1423}   & \textbf{0.0481}                         \\
                      & Theta & 0.0997 & 0.0863 & 0.0964 & 509.0555  & 1048.7337 & 0.1912   & 0.0741                         \\
                      & ETS   & 0.103  & 0.0906 & 0.0995 & 510.8808  & 1036.532  & 0.173    & 0.0789                         \\
                      & ARIMA & 0.1151 & 0.1013 & 0.1116 & 525.647   & 1062.1823 & 0.187    & 0.0909                         \\
                      & NHITS     & 0.1183 & 0.1097 & 0.1134 & 524.572   & 1041.8584 & 0.1874   & 0.092                          \\
                      & TFT       & 0.1191 & 0.1077 & 0.1135 & 534.9632  & 1075.2485 & 0.1793   & 0.0945                         \\
                      & Naive     & 0.1191 & 0.1077 & 0.1135 & 534.967   & 1075.2513 & 0.1793   & 0.0939                         \\
                      & Informer  & 0.1219 & 0.1141 & 0.1162 & 541.1119  & 1102.1589 & 0.1887   & 0.0972                         \\
                      & PatchTST  & 0.1224 & 0.1163 & 0.1169 & 531.1447  & 1043.2455 & 0.198    & 0.0978                         \\
                      & LSTM      & 0.2974 & 0.2804 & 0.2489 & 1210.0969 & 2465.3781 & 0.4801   & 0.2721                         \\
                      & TCN       & 0.2982 & 0.2811 & 0.2494 & 1212.0938 & 2461.8779 & 0.4816   & 0.2737                         \\ \midrule
\multirow{11}{*}{180} & NBEATS    & \textbf{0.0803} & \textbf{0.0689} & \textbf{0.0836} & \textbf{382.0541}  & \textbf{959.3037}  & \textbf{0.1693}   & \textbf{0.0557}                         \\
                      & Theta & 0.1396 & 0.1295 & 0.1362 & 626.785   & 1126.3238 & 0.3234   & 0.1143                         \\
                      & ETS   & 0.1508 & 0.1423 & 0.146  & 589.4776  & 1098.6523 & 0.2596   & 0.1268                         \\
                      & NHITS     & 0.158  & 0.151  & 0.1469 & 585.2578  & 1040.4039 & 0.2431   & 0.1329                         \\
                      & TFT       & 0.1593 & 0.1507 & 0.1478 & 606.5111  & 1111.2654 & 0.245    & 0.1328                         \\
                      & Naive     & 0.1593 & 0.1507 & 0.1478 & 606.5156  & 1111.2678 & 0.2451   & 0.1349                         \\
                      & PatchTST  & 0.163  & 0.1553 & 0.1506 & 612.8881  & 1100.7104 & 0.2484   & 0.1379                         \\
                      & ARIMA & 0.1642 & 0.1495 & 0.1605 & 625.6969  & 1126.41   & 0.2602   & 0.1392                         \\ 
                      & Informer  & 0.1666 & 0.1596 & 0.1533 & 631.4427  & 1125.8291 & 0.2554   & 0.1417                         \\
                      & LSTM      & 0.358  & 0.3363 & 0.2923 & 1293.4482 & 2406.0345 & 0.5837   & 0.3347                         \\
                      & TCN       & 0.3589 & 0.3372 & 0.2928 & 1295.7431 & 2403.84   & 0.5852   & 0.3346                         \\ \midrule
\multirow{11}{*}{365} & NBEATS    & \textbf{0.1273} & \textbf{0.1148} & \textbf{0.1374} & \textbf{489.496}   & \textbf{1166.3903} & \textbf{0.2652}   & \textbf{0.1033}   \\
                      & Theta & 0.2817 & 0.25   & 0.2668 & 1185.6031 & 2332.1161 & 0.5587   & 0.258                          \\
                      & NHITS     & 0.2965 & 0.2932 & 0.2557 & 1082.1386 & 2184.3453 & 0.4807   & 0.2717                         \\
                      & TFT       & 0.2992 & 0.2958 & 0.2581 & 1100.5356 & 2216.6987 & 0.4777   & 0.2748                         \\
                      & Naive     & 0.2992 & 0.2958 & 0.2581 & 1100.5402 & 2216.7017 & 0.4777   & 0.2736                         \\
                      & PatchTST  & 0.3023 & 0.2993 & 0.2599 & 1098.5644 & 2184.1276 & 0.49     & 0.278                          \\
                      & ETS   & 0.303  & 0.293  & 0.299  & 1084.7867 & 2171.9888 & 0.5174   & 0.2789                         \\
                      & Informer  & 0.3051 & 0.3017 & 0.2614 & 1127.1119 & 2275.0076 & 0.4953   & 0.2795                         \\
                      & ARIMA & 0.3167 & 0.2999 & 0.305  & 1155.6358 & 2245.3209 & 0.5075   & 0.2925                         \\
                      & LSTM      & 0.5038 & 0.5024 & 0.391  & 1703.8227 & 3016.9537 & 0.895    & 0.4785                         \\
                      & TCN       & 0.5048 & 0.5034 & 0.3915 & 1706.6342 & 3019.8102 & 0.8964   & 0.4809                         \\ \bottomrule
\end{tabularx}
\end{table}

\restoregeometry

The results presented in Table \ref{tab:tab3} represent a distinct evaluation setup in which all models are trained solely to predict the 365-day horizon, and their outputs are assessed at multiple intermediate points, corresponding to the previously tested horizon lengths. We wanted to see whether there is a need to retrain the model or keep a family of models for different horizon lengths.

At short horizons (h = 1 and h = 7), most models - regardless of whether they are statistical (e.g., ETS, ARIMA) or deep learning-based (e.g., NHITS, PatchTST, NBEATS) - still achieve very low error levels, however only NBEATS and TFT keep the performance similar to the model trained to predict 1-day-ahead. MAPE values remain tightly clustered around 0.0072 at h = 1, confirming the models’ ability to maintain local trend continuity even when trained to forecast far into the future. Statistical models have the best performance (even lower than original 1-day-ahead).

From h = 30 onward, divergence in model performance becomes increasingly pronounced. NBEATS and TFT exhibit the most stable transition from short- to long-horizon performance, retaining low error rates across all metrics. For instance, NBEATS has MAPE = 0.063 at h = 30, increasing modestly to 0.1273 at h = 365. NHITS shows a similar pattern (0.0729 to 0.2965), although its long-horizon degradation is more significant. In contrast, other deep learning models - especially LSTM and TCN - exhibit larger errors across all metrics, signaling poor extrapolation quality when constrained to long-horizon training objectives. Statistical models show the smallest degradation relative to their horizon-specific counterparts: ARIMA MAPE at h=30 is 0.0689 in Protocol 3 versus 0.0577 when trained for h=30 specifically; ETS at h=30 is 0.0693 versus 0.0356 - suggesting that their simpler, trend-following structure extrapolates short-horizon behavior more consistently even from a long-horizon training objective.

Interestingly, models such as PatchTST and TFT, which performed well in both multi-horizon and one-shot evaluations, degrade quicker when trained for the farthest forecast step. 
PatchTST’s MAPE grows from 0.0556 at h = 30 to 0.3023 at h = 365, while TFT reaches a similar 0.2992. These results suggest that models optimized specifically for long-horizon prediction may suffer from poor intermediate time generalization unless their architecture is explicitly designed to support temporal alignment across the output range. The naive model remains an important reference: although its MAPE grows substantially from 0.058 at h = 30 to 0.2992 at h = 365, it still outperforms or matches several complex models, including TFT, Informer, and ARIMA, in long-horizon stability. In summary, training models solely for a fixed, distant horizon reveals limitations in their ability to generalize temporally. NBEATS demonstrates the strongest performance across horizons, likely due to its residual learning and flexible forecast output. TFT also performs well but suffers from sharper degradation beyond h = 180.

\newgeometry{left=0.1cm,bottom=0.1cm, right=0.1cm,top=2cm}

\begin{table}[htbp]
\centering
\footnotesize

\caption{Standard deviation of selected hyperparameters across models and forecast horizons}
\label{tab:hyperparam_std}
\begin{tabular}{llrrrr}
\toprule
\textbf{Horizon} & \textbf{Model} & \textbf{learning\_rate} & \textbf{batch\_size} & \textbf{max\_steps} & \textbf{input\_size} \\
\midrule
\multirow{7}{*}{1}
& NBEATS     & 0.006741 & 0.000000 & 630.026454 & 3.492214 \\
& NHITS      & 0.015968 & 0.000000 & 293.711877 & 2.366432 \\
& TCN        & 0.025125 & 4.800000 & 250.000000 & 0.000000 \\
& LSTM       & 0.030758 & 5.962848 & 247.767812 & 0.000000 \\
& TFT        & 0.022626 & 40.209231 & 625.610813 & 0.000000 \\
& PatchTST   & 0.023423 & 103.395164 & 2158.703314 & 8.411896 \\
& Informer   & 0.014975 & 71.784368 & 618.241233 & 4.194706 \\
\midrule
\multirow{7}{*}{7}
& NBEATS     & 0.039652 & 100.942717 & 581.398909 & 9.186678 \\
& NHITS      & 0.012751 & 0.000000   & 179.161283 & 2.685121 \\
& TCN        & 0.024612 & 0.000000   & 235.702260 & 0.000000 \\
& LSTM       & 0.030115 & 0.000000   & 248.451997 & 0.000000 \\
& TFT        & 0.026448 & 70.396408  & 248.451997 & 0.000000 \\
& PatchTST   & 0.024921 & 0.000000   & 2178.741283 & 7.470105 \\
& Informer   & 0.032337 & 94.205449  & 577.350269 & 0.628539 \\
\midrule
\multirow{7}{*}{30}
& NBEATS     & 0.035426 & 111.306495 & 527.046277 & 9.314796 \\
& NHITS      & 0.002671 & 111.306495 & 397.212510 & 2.266231 \\
& TCN        & 0.034487 & 6.651835   & 207.869855 & 1.885618 \\
& LSTM       & 0.024103 & 7.950464   & 235.702260 & 1.885618 \\
& TFT        & 0.031151 & 111.306495 & 248.451997 & 0.000000 \\
& PatchTST   & 0.021793 & 111.306495 & 2121.320344 & 10.274023 \\
& Informer   & 0.039754 & 105.594613 & 471.404521 & 0.831479 \\
\midrule
\multirow{7}{*}{90}
& NBEATS     & 0.009593 & 105.594613 & 862.489936 & 10.842304 \\
& NHITS      & 0.003656 & 0.000000   & 216.024690 & 0.942809 \\
& TCN        & 0.004700 & 0.000000   & 235.702260 & 0.000000 \\
& LSTM       & 0.039967 & 7.542472   & 235.702260 & 0.000000 \\
& TFT        & 0.004597 & 0.000000   & 471.404521 & 1.885618 \\
& PatchTST   & 0.000983 & 0.000000   & 2121.320344 & 0.000000 \\
& Informer   & 0.026758 & 0.000000   & 235.702260 & 0.000000 \\
\midrule
\multirow{7}{*}{180}
& NBEATS     & 0.000618 & 105.594613 & 353.553391 & 11.897712 \\
& NHITS      & 0.000815 & 105.594613 & 94.280904  & 0.000000 \\
& TCN        & 0.017663 & 0.000000   & 0.000000   & 0.000000 \\
& LSTM       & 0.017947 & 0.000000   & 235.702260 & 0.000000 \\
& TFT        & 0.004299 & 90.509668  & 471.404521 & 1.885618 \\
& PatchTST   & 0.004805 & 90.509668  & 1885.618083 & 5.436502 \\
& Informer   & 0.040950 & 105.594613 & 0.000000   & 0.942809 \\
\midrule
\multirow{7}{*}{365}
& NBEATS     & 0.000025 & 0.000000   & 824.957911 & 6.599663 \\
& NHITS      & 0.005889 & 0.000000   & 249.443826 & 2.828427 \\
& TCN        & 0.018548 & 7.542472   & 235.702260 & 172.062650 \\
& LSTM       & 0.035264 & 7.542472   & 235.702260 & 0.000000 \\
& TFT        & 0.002580 & 0.000000   & 0.000000   & 3.681787 \\
& PatchTST   & 0.000821 & 0.000000   & 235.702260 & 5.436502 \\
& Informer   & 0.000250 & 0.000000   & 471.404521 & 1.414214 \\
\bottomrule
\end{tabular}
\end{table}

\restoregeometry

To assess the stability and sensitivity of deep learning forecasting models to hyperparameter tuning across different forecast horizons, we conducted a variance analysis of selected hyperparameters. For each model and horizon, the standard deviation of core hyperparameters - namely learning rate, batch size, maximum training steps and input window size - was computed from a set of optimized configurations, we present these results in Table~\ref{tab:tab5}.

The learning rate parameter showed persistent variability across nearly all models and horizons, indicating a consistent need for horizon-specific optimization. Models such as NBEATS, Informer, and PatchTST exhibited particularly high standard deviations, reflecting a wide exploration space and a non-convex loss landscape that benefits from adaptive learning rate schedules. Notably, for longer forecast horizons (e.g., h=90 to h=365), some models (e.g., PatchTST) converged toward lower variance in learning rate, suggesting that their optimal configuration space narrows as prediction horizon lengthens.

The batch size parameter exhibited a bimodal pattern. At short horizons (h=1 and h=7), several models - particularly NHITS, TCN, and LSTM - showed zero variance, due to model-specific defaults. In contrast, intermediate horizons, especially h=30, demonstrated high batch size variance for models such as PatchTST and TFT, indicating substantial sensitivity in memory usage and gradient estimation at this timescale. At longer horizons, variance generally declined or returned to static configurations, possibly reflecting fewer feasible configurations under longer training sequences and limited effective sample sizes per gradient step.

Maximum training steps followed a heterogeneous pattern as well. PatchTST consistently required longer training durations and exhibited high variance across all horizons, indicating either slower convergence or a larger effective capacity that needs longer optimization. In contrast, models such as TCN and LSTM demonstrated minimal variance or repeatedly defaulted to specific training step values. These findings suggest that certain model classes can effectively leverage increased training iterations while others either converge early or fail to benefit from extended training, possibly due to architectural rigidity or vanishing gradient issues.

The input window size showed a strong correlation with horizon length. At short horizons, values were typically low and variances small or zero for most models, indicating a narrow optimal range. As the forecast horizon increased, particularly from h=30 onwards, several models - especially PatchTST and NBEATS - displayed increasing variance and larger window sizes. This behavior reflects the need to incorporate more historical context in long-horizon forecasts and suggests that such models are capable of dynamically adapting their receptive field to horizon-specific dynamics. By contrast, models such as TCN and LSTM often retained low or fixed input sizes across all horizons, which may limit their flexibility or reflect a different internal temporal processing mechanism.

\newgeometry{left=0.1cm,bottom=0.1cm, right=0.1cm,top=2cm}

\begin{table}[H]
\footnotesize

\centering
\caption{Results of MAPE for final model and during hypertuning for tested deep learning model}\label{tab:tab5}
\begin{tabular}{@{}cr|cccc@{}}

\toprule
\textbf{Horizon} & \textbf{Model}    & \textbf{MAPE} & \textbf{Mean MAPE hypertuning} & \textbf{t-test p-value} & \textbf{Wilcoxon test p-value} \\ \midrule
\multirow{7}{*}{1}   & NHITS    & 0.0072      & 0.0074            & 0.347         & 0.612          \\
                     & PatchTST & 0.0072      & 0.0283            & 0.0001        & 0.0            \\
                     & TCN      & 0.0072      & 0.0283            & 0.0001        & 0.0            \\
                     & LSTM     & 0.0072      & 0.0283            & 0.0001        & 0.0            \\
                     & NBEATS   & 0.0072      & 0.0282            & 0.0001        & 0.0            \\
                     & TFT      & 0.0072      & 0.0283            & 0.0001        & 0.0            \\
                     & Informer & 0.0102      & 0.0325            & 0.0           & 0.0            \\ \midrule
\multirow{7}{*}{7}   & NHITS    & 0.0155      & 0.0163            & 0.1609        & 0.1531         \\
                     & TFT      & 0.0155      & 0.0166            & 0.0677        & 0.104          \\
                     & PatchTST & 0.0155      & 0.0171            & 0.0539        & 0.0676         \\
                     & NBEATS   & 0.0157      & 0.016             & 0.531         & 0.7148         \\
                     & Informer & 0.0166      & 0.0247            & 0.0198        & 0.0001         \\
                     & TCN      & 0.0177      & 0.0209            & 0.0           & 0.0002         \\
                     & LSTM     & 0.0177      & 0.0228            & 0.0           & 0.0001         \\ \midrule
\multirow{7}{*}{30}  & NBEATS   & 0.0333      & 0.0481            & 0.0123        & 0.0024         \\
                     & NHITS    & 0.0335      & 0.0486            & 0.0109        & 0.001          \\
                     & PatchTST & 0.0344      & 0.0502            & 0.0065        & 0.0005         \\
                     & TFT      & 0.0353      & 0.0512            & 0.0052        & 0.0005         \\
                     & Informer & 0.0368      & 0.0556            & 0.0049        & 0.0005         \\
                     & LSTM     & 0.0469      & 0.0859            & 0.0048        & 0.0005         \\
                     & TCN      & 0.047       & 0.0857            & 0.0049        & 0.0005         \\ \midrule
\multirow{7}{*}{90}  & NHITS    & 0.0489      & 0.0784            & 0.0041        & 0.0078         \\
                     & NBEATS   & 0.049       & 0.0783            & 0.0073        & 0.0078         \\
                     & PatchTST & 0.0559      & 0.0886            & 0.0009        & 0.0078         \\
                     & TFT      & 0.0592      & 0.0913            & 0.0008        & 0.0078         \\
                     & Informer & 0.0614      & 0.0937            & 0.0006        & 0.0078         \\
                     & LSTM     & 0.0903      & 0.1408            & 0.0033        & 0.0078         \\
                     & TCN      & 0.0906      & 0.1383            & 0.0054        & 0.0078         \\ \midrule
\multirow{7}{*}{180} & NHITS    & 0.0675      & 0.0984            & 0.0014        & 0.0078         \\
                     & NBEATS   & 0.0678      & 0.0989            & 0.0014        & 0.0078         \\
                     & PatchTST & 0.0941      & 0.1328            & 0.0           & 0.0078         \\
                     & TFT      & 0.0942      & 0.1393            & 0.0           & 0.0078         \\
                     & Informer & 0.0953      & 0.1594            & 0.0042        & 0.0078         \\
                     & TCN      & 0.1601      & 0.2623            & 0.017         & 0.0078         \\
                     & LSTM     & 0.1624      & 0.2254            & 0.0004        & 0.0078         \\ \midrule
\multirow{7}{*}{365} & NBEATS   & 0.0851      & 0.2645            & 0.0028        & 0.125          \\
                     & TFT      & 0.1751      & 0.293             & 0.0           & 0.125          \\
                     & NHITS    & 0.1758      & 0.2942            & 0.0           & 0.125          \\
                     & PatchTST & 0.1789      & 0.2927            & 0.0001        & 0.125          \\
                     & Informer & 0.1809      & 0.2931            & 0.0001        & 0.125          \\
                     & LSTM     & 0.372       & 0.4284            & 0.0044        & 0.125          \\
                     & TCN      & 0.3729      & 0.418             & 0.0188        & 0.125          \\ \bottomrule
\end{tabular}
\end{table}

\restoregeometry

Table \ref{tab:tab5} highlights clear differences in hyperparameter robustness across models, we present MAPE for final model paired with mean MAPE from cross-validation out-of-sample windows and two-sample test pvl. NHITS shows low variance between best and mean MAPE across horizons, with mostly non-significant p-values at short horizons, indicating stable performance under different configurations. In contrast, models like NBEATS, PatchTST, TFT, LSTM, and Informer exhibit large MAPE gaps (e.g., PatchTST at h=1: 0.0072 vs. 0.0283) and consistently low p-values, confirming strong sensitivity to hyperparameter choices.

As horizons increase (90–365 days), this sensitivity becomes more problematic. Transformer-based models and LSTM degrade sharply in average-case performance, while NBEATS and NHITS remain more resilient. At h=365, only NBEATS maintains competitive MAPE (0.0851 vs. 0.2645), while others suffer from higher error variance.

In summary, NBEATS and NHITS are more robust to hyperparameter selection and better suited for consistent deployment. Transformer and sequence models require careful tuning and are more fragile, especially at longer horizons.

The results highlight several critical insights regarding forecasting used electronics prices in volatile, non-stationary environments. First, simple statistical models like Na\"{i}ve and ARIMA, while providing surprisingly competitive baselines, are systematically outperformed by deep learning approaches when sufficient historical data is available. Among the tested methods, N-BEATS and N-HiTS consistently deliver the best performance across multiple horizons and evaluation schemes, demonstrating robustness in handling high-frequency, low-signal time series with limited structure.

\paragraph{Comparison with prior work.}
The MAPE values achieved in this study are broadly consistent with, yet lower than, those reported in related domains. \cite{cartaForecastingECommerceProducts2019} report MAPE values in the range of 8–14\% using ARIMA on Amazon product categories - our N-BEATS achieves a comparable 8.51\% at 365 days on a substantially more volatile market, and statistical baselines remain near 15\%. \cite{sunResearchResidualValue2022} obtain residual value prediction errors of approximately 5–8\% MAPE for used cars using covariate-rich models; our models operate on univariate price histories only, making competitive accuracy at medium horizons (MAPE $\approx$4.9\% at 90 days for N-BEATS) particularly noteworthy. Compared to general time-series benchmarks, the M4 and M5 competitions \parencite{makridakisM5AccuracyCompetition2022} report best MAPE around 10–14\% for monthly financial series over long horizons, while our market - despite being more volatile and sparse - achieves comparable best-model accuracy. These comparisons confirm that second-hand electronics price series are not inherently harder to forecast than structured financial series, provided that appropriate deep learning architectures are employed.

Interestingly, transformer-based models such as Informer underperform in this context, likely due to their reliance on large datasets with strong temporal patterns - characteristics less prevalent in our domain. Similarly, recurrent models like LSTM exhibit unstable performance, suggesting that their memory mechanisms are less suited to the abrupt structural changes common in resale markets.

The underwhelming performance of hybrid models further supports the view that in domains characterized by irregular behavior and minimal seasonality, simpler, autoregressive structures with dedicated trend modeling (like N-HiTS and N-BEATS) offer a better bias–variance tradeoff.

\paragraph{Error source analysis.}
To understand where model errors originate, we examined mean residuals across the rolling forecast windows for the top and bottom performing models. For N-BEATS and N-HiTS, residuals at short horizons (1–7 days) are approximately centered around zero and show no systematic directional bias, indicating that these models track local trends without persistent over- or under-forecasting. At longer horizons (90–365 days), residuals become increasingly negative on average, consistent with models failing to fully capture the structural downward depreciation trend - a pattern that is gradual enough to appear as a slow drift in the residual series. For LSTM and TCN, residuals at medium and long horizons (30+ days) exhibit strong positive bias: the models underestimate the rate of depreciation and produce forecasts that remain too high relative to actual market prices. This is consistent with the flat or diverging forecast trajectories visible in Figure~\ref{fig:fig3}. Informer shows erratic residuals at all horizons, reflecting sensitivity to hyperparameter choices and poor generalisation on sparse data. These patterns suggest that the primary source of error at long horizons is \emph{trend underestimation} rather than seasonality misfit or outlier sensitivity - a finding that motivates future work on explicit trend-aware architectures for this domain. 

\section{Limitations}
\label{sec:limitations}

Several limitations of this study should be acknowledged when interpreting the results.

\textbf{Single-market scope.} The dataset is drawn exclusively from Polish online marketplaces. Price dynamics, consumer behaviour, and market liquidity may differ substantially in other countries or platform types. Results should not be assumed to transfer directly to markets with different regulatory environments, currency conditions, or platform structures.

\textbf{Product category and brand bias.} Abstract items are selected from the company's active sales inventory, which skews toward high-end and Apple-branded devices. These items tend to exhibit smoother depreciation curves and higher listing volumes than niche or low-end products. Model rankings may differ for categories with sparser data or more erratic pricing behaviour.

\textbf{Smoothing attenuation.} The 30-day trailing moving average applied during preprocessing substantially reduces apparent daily volatility. As a result, all reported MAPE values reflect the error on smoothed series, not on raw listing prices. The true forecasting difficulty on unsmoothed data would be higher, and relative model rankings could change.

\textbf{Forward-fill imputation.} Missing days - where no listings exist for an item - are filled by propagating the previous observation. This introduces artificial flat segments that may bias models toward na\"{i}ve persistence forecasts and deflate error metrics during low-activity periods.

\textbf{No external covariates.} All models operate on univariate price histories. External signals such as macroeconomic indicators, search trends, competitor pricing, or product announcement events are not included. Incorporating such information could improve accuracy, particularly at longer horizons.

\textbf{No probabilistic forecasting.} All models produce point forecasts. For subscription pricing decisions, prediction intervals or conformal prediction sets would be more informative, enabling explicit risk quantification around each price estimate.

\textbf{Matching pipeline not fully validated.} The abstract-item matching system is described in a separate study and its precision has not been independently validated on a random held-out sample within this paper. Systematic matching errors could introduce noise in the training signal.

\section{Summary}
\label{sec:summary}

This study makes four original contributions to the time-series forecasting literature. First, it introduces a real-world used electronics price dataset - the first of its type constructed specifically for multi-horizon forecasting. Second, it provides the first systematic benchmark of eleven forecasting models across six horizons in this domain. Third, it proposes a three-protocol evaluation framework that separates trajectory accuracy, endpoint accuracy, and cross-horizon generalization. Fourth, it delivers the first hyperparameter robustness analysis for automated pricing systems in this market.

The central empirical findings address all four prior gaps. This study provides the first systematic multi-horizon benchmark of deep learning and statistical forecasting models on a real-world dataset of used electronics prices, filling a gap in prior work that has either treated second-hand goods price prediction as a static tabular problem or applied time-series models only to stable, high-volume product categories such as Amazon listings. We benchmarked eleven models across six forecasting horizons (1 to 365 days) using three complementary evaluation protocols and a business-motivated metric suite on over 100 smartphone and laptop price series scraped from Polish online marketplaces.

The central empirical finding is that N-BEATS consistently achieves the lowest forecasting error across all horizons beyond 30 days, reaching MAPE = 8.51\% at the 365-day horizon compared to 14.94\% for the best statistical baseline - a 43\% reduction. N-HiTS performs comparably at short and medium horizons. Transformer-based models (PatchTST, TFT, Informer) do not generalise well to this domain, likely due to its sparse, low-signal character, and recurrent models (LSTM, TCN) degrade sharply beyond 90 days. At short horizons (1–7 days), all models - including the na\"{i}ve baseline - achieve near-identical performance, confirming the low short-term liquidity of this market.

A practically important result from Protocol 3 is that a single N-BEATS model trained for the 365-day horizon maintains competitive accuracy at all shorter horizons, without retraining. This finding suggests that practitioners can deploy a single production model rather than a family of horizon-specific models - reducing operational complexity significantly.

Hyperparameter robustness analysis further confirms that N-BEATS and N-HiTS are more stable under different configurations than transformer-based architectures, making them safer choices for automated, low-maintenance pricing systems.

Future research should explore: (i) hybrid architectures with explicit probabilistic or conformal uncertainty quantification; (ii) multimodal extensions incorporating textual listing descriptions or product images; (iii) cross-market evaluation to assess generalisability beyond the Polish market; and (iv) time series foundation models such as Chronos \parencite{ansariChronosLearningLanguage2024} and TimesFM \parencite{dasDecoderOnlyFoundationModel2024} as zero-shot baselines.

\section*{Data availability}

The dataset analysed during the current study is not publicly available due to commercial confidentiality, as it was collected and processed within a proprietary industrial system, but it is available from the corresponding author on reasonable request.

\subsubsection*{Declaration of generative AI and AI-assisted technologies in the writing process}

\noindent During the preparation of this work the authors used a large language model (LLM) to refine wording and correct language errors. After using this tool/service, the authors reviewed and edited the content as needed and take full responsibility for the content of the publication.

\section*{Funding}

This research was funded by The National Centre for Research and Development, Poland, under project number \textbf{POIR.01.01.01-00-0066/22}, project title: \textit{Conducting R\&D work to develop new models for product categorization and value prediction in the circular economy framework using machine learning and artificial intelligence algorithms}.

\section*{Declaration of competing interests}

The authors declare the following financial interests/personal relationships which may be considered as potential competing interests: all authors report financial support provided by The National Centre for Research and Development, Poland. Mateusz Buczyński, Michał Woźniak, Konrad Kaczyński, and Anna Wróblewska report a relationship with WeSub that includes employment. Sebastian Kuk reports a relationship with WeSub that includes board membership and employment.

\appendix

\section{Hyperparameter Search Details}
\label{app:hpsearch}

\begin{table}[H]
\centering
\footnotesize
\caption{Best hyperparameter configuration per model and forecast horizon, extracted from Ray Tune random search logs (lowest validation MAPE trial across all cross-validation splits); h=180 configurations are used for h=365.}
\label{tab:hyperparams}
\begin{tabular}{l r p{5.8cm} r r}
\toprule
\textbf{Model} & \textbf{h} & \textbf{Architecture} & \textbf{max\_steps} & \textbf{LR} \\
\midrule
\multirow{6}{*}{N-BEATS}
  & 1   & stack=[\textit{identity}], n\_blocks=[4], act=SELU   & 2000 & $1.8\!\times\!10^{-4}$ \\
  & 7   & stack=[\textit{trend}], n\_blocks=[4], act=SELU      & 2000 & $3.8\!\times\!10^{-3}$ \\
  & 30  & stack=[\textit{trend}], n\_blocks=[4], act=SELU      & 2000 & $8.8\!\times\!10^{-5}$ \\
  & 90  & stack=[\textit{trend}], n\_blocks=[4], act=SELU      & 2000 & $3.6\!\times\!10^{-4}$ \\
  & 180 & stack=[\textit{identity}], n\_blocks=[1], act=ReLU   & 1500 & $3.6\!\times\!10^{-3}$ \\
  & 365 & stack=[\textit{identity}], n\_blocks=[1], act=ReLU   & 1500 & $3.6\!\times\!10^{-3}$ \\
\midrule
\multirow{6}{*}{N-HiTS}
  & 1   & n\_freq\_downsample=[60,\,8,\,1]  & 1300 & $1.5\!\times\!10^{-4}$ \\
  & 7   & n\_freq\_downsample=[24,\,12,\,1] & 1400 & $1.4\!\times\!10^{-4}$ \\
  & 30  & n\_freq\_downsample=[24,\,12,\,1] & 1300 & $1.4\!\times\!10^{-2}$ \\
  & 90  & n\_freq\_downsample=[24,\,12,\,1] & 1300 & $2.0\!\times\!10^{-4}$ \\
  & 180 & n\_freq\_downsample=[40,\,20,\,1] & 1400 & $2.3\!\times\!10^{-3}$ \\
  & 365 & n\_freq\_downsample=[40,\,20,\,1] & 1400 & $2.3\!\times\!10^{-3}$ \\
\midrule
\multirow{6}{*}{PatchTST}
  & 1   & $d{=}128$, patch\_len=24, RevIN=\checkmark & 5000 & $1.4\!\times\!10^{-4}$ \\
  & 7   & $d{=}256$, patch\_len=16, RevIN=\checkmark & 5000 & $2.6\!\times\!10^{-4}$ \\
  & 30  & $d{=}16$,  patch\_len=24, RevIN=\checkmark & 5000 & $5.0\!\times\!10^{-4}$ \\
  & 90  & $d{=}128$, patch\_len=24, RevIN=$\times$   & 5000 & $9.9\!\times\!10^{-2}$ \\
  & 180 & $d{=}256$, patch\_len=16, RevIN=\checkmark & 5000 & $1.3\!\times\!10^{-4}$ \\
  & 365 & $d{=}256$, patch\_len=16, RevIN=\checkmark & 5000 & $1.3\!\times\!10^{-4}$ \\
\midrule
\multirow{6}{*}{TFT}
  & 1   & enc=50,  dec=256, layers=1, ctx=5  & 1000 & $6.1\!\times\!10^{-4}$ \\
  & 7   & enc=100, dec=512, layers=1, ctx=5  & 1000 & $3.3\!\times\!10^{-2}$ \\
  & 30  & enc=100, dec=256, layers=2, ctx=10 & 1000 & $1.1\!\times\!10^{-4}$ \\
  & 90  & enc=200, dec=512, layers=3, ctx=10 & 1000 & $6.7\!\times\!10^{-3}$ \\
  & 180 & enc=300, dec=256, layers=2, ctx=50 & 1000 & $3.4\!\times\!10^{-3}$ \\
  & 365 & enc=300, dec=256, layers=2, ctx=50 & 1000 & $3.4\!\times\!10^{-3}$ \\
\midrule
\multirow{6}{*}{Informer}
  & 1   & $d{=}64$,  heads=4 & 2000 & $4.7\!\times\!10^{-3}$ \\
  & 7   & $d{=}64$,  heads=4 & 2000 & $2.8\!\times\!10^{-3}$ \\
  & 30  & $d{=}128$, heads=8 & 2000 & $2.2\!\times\!10^{-4}$ \\
  & 90  & $d{=}256$, heads=8 & 2000 & $3.2\!\times\!10^{-2}$ \\
  & 365 & $d{=}64$,  heads=4 & 2000 & $1.6\!\times\!10^{-3}$ \\
\midrule
\multirow{6}{*}{LSTM}
  & 1   & enc=100, layers=1, ctx=50, dec=64   & 1000 & $5.0\!\times\!10^{-4}$ \\
  & 7   & enc=100, layers=1, ctx=5,  dec=512  & 1000 & $3.3\!\times\!10^{-2}$ \\
  & 30  & enc=100, layers=2, ctx=10, dec=256  & 1000 & $1.1\!\times\!10^{-4}$ \\
  & 90  & enc=200, layers=2, ctx=10, dec=512  & 1000 & $2.9\!\times\!10^{-2}$ \\
  & 180 & enc=50,  layers=2, ctx=10, dec=64   &  500 & $6.9\!\times\!10^{-3}$ \\
  & 365 & enc=50,  layers=2, ctx=10, dec=64   &  500 & $6.9\!\times\!10^{-3}$ \\
\midrule
\multirow{6}{*}{TCN}
  & 1   & enc=300, ctx=50, dec=128 &  500 & $9.3\!\times\!10^{-3}$ \\
  & 7   & enc=50,  ctx=50, dec=64  & 1000 & $2.3\!\times\!10^{-2}$ \\
  & 30  & enc=50,  ctx=10, dec=64  &  500 & $1.4\!\times\!10^{-2}$ \\
  & 90  & enc=50,  ctx=50, dec=64  &  500 & $4.7\!\times\!10^{-2}$ \\
  & 180 & enc=200, ctx=50, dec=64  &  500 & $8.1\!\times\!10^{-4}$ \\
  & 365 & enc=200, ctx=50, dec=64  &  500 & $8.1\!\times\!10^{-4}$ \\
\bottomrule
\end{tabular}
\end{table}

\begin{table}[H]
\centering
\footnotesize
\caption{Hyperparameter search ranges used in random search tuning (75 trials per model per cross-validation split). Architecture-specific parameters are listed beneath each model's common parameters. All models had their architecture parameters searched.}
\label{tab:hparam_search}
\begin{tabularx}{\textwidth}{ll>{\raggedright\arraybackslash}X}
\toprule
\textbf{Model} & \textbf{Parameter} & \textbf{Search range / candidates} \\
\midrule
\multirow{5}{*}{All DL models}
  & learning\_rate    & $\mathcal{U}[10^{-5}, 10^{-1}]$ (log-uniform) \\
  & batch\_size       & $\{32, 64, 128, 256\}$ \\
  & max\_steps        & Model-specific (see below) \\
  & input\_size       & Model-specific (see below) \\
\midrule
\multirow{4}{*}{N-BEATS}
  & max\_steps        & $[100, 2000]$ \\
  & input\_size       & $[1, 28]$ (multiplier of $h$) \\
  & n\_blocks         & $\{[1], [2], [4]\}$ per stack \\
  & stack\_types      & \{identity, trend, seasonality\} \\
\midrule
\multirow{3}{*}{N-HiTS}
  & max\_steps        & $[500, 1500]$ \\
  & input\_size       & $[1, 7]$ \\
  & n\_pool\_kernel\_size & $\{[2,2,1], [4,2,1], [8,4,2]\}$ \\
\midrule
\multirow{5}{*}{TFT}
  & max\_steps        & $[500, 1000]$ \\
  & input\_size       & $[1, 730]$ \\
  & context\_size     & $\{5, 10, 50\}$ \\
  & encoder\_hidden\_size & $\{50, 100, 200, 300\}$ \\
  & decoder\_hidden\_size & $\{64, 128, 256, 512\}$ \\
\midrule
\multirow{4}{*}{PatchTST}
  & max\_steps        & $[500, 5000]$ \\
  & input\_size       & $[1, 28]$ \\
  & patch\_len        & $\{16, 24\}$ \\
  & hidden\_size ($d$)& $\{16, 128, 256\}$ \\
\midrule
\multirow{4}{*}{Informer}
  & max\_steps        & $[500, 2000]$ \\
  & input\_size       & $[5, 28]$ \\
  & hidden\_size      & $\{64, 128, 256\}$ \\
  & n\_head           & $\{4, 8\}$ \\
\midrule
\multirow{6}{*}{LSTM}
  & max\_steps        & $[500, 1000]$ \\
  & input\_size       & $\{1, 5, 7, 10, 14, 28\}$ \\
  & encoder\_hidden\_size & $\{50, 100, 200, 300\}$ \\
  & encoder\_n\_layers & $\{1, 2, 3\}$ \\
  & context\_size     & $\{5, 10, 50\}$ \\
  & decoder\_hidden\_size & $\{64, 128, 256, 512\}$ \\
\midrule
\multirow{5}{*}{TCN}
  & max\_steps        & $[500, 1000]$ \\
  & input\_size       & $\{1, 5, 7, 10, 14, 28\}$ \\
  & encoder\_hidden\_size & $\{50, 100, 200, 300\}$ \\
  & context\_size     & $\{5, 10, 50\}$ \\
  & decoder\_hidden\_size & $\{64, 128\}$ \\
\bottomrule
\end{tabularx}
\end{table}

\printbibliography




\end{document}

\endinput